\documentclass[letterpaper,twocolumn,10pt]{article}

\usepackage{usenix}

\usepackage[square,comma,numbers,sort&compress]{natbib}

\usepackage{amsmath,amsopn}
\usepackage{xspace}
\usepackage{graphicx}
\usepackage{fancyvrb}
\usepackage{multirow}
\usepackage{array}
\usepackage{underscore}
\usepackage{relsize}
\usepackage{pbox}
\usepackage{xstring}
\usepackage{tikz}
\usetikzlibrary{arrows.meta,positioning,backgrounds,fit,calc}
\usepackage{fp}
\usepackage{flushend}
\usepackage{balance}
\usepackage{booktabs}
\usepackage{makecell}
\usepackage{enumitem}
\usepackage{float}
\usepackage{capt-of}

\usepackage{color, colortbl}


\makeatletter
\def\mdseries@tt{m}
\makeatother
\usepackage{minted}

\fvset{fontsize=\scriptsize,xleftmargin=14pt,numbers=left,numbersep=5pt}

\makeatletter
\def\PY@reset{\let\PY@it=\relax \let\PY@bf=\relax%
    \let\PY@ul=\relax \let\PY@tc=\relax%
    \let\PY@bc=\relax \let\PY@ff=\relax}
\def\PY@tok#1{\csname PY@tok@#1\endcsname}
\def\PY@toks#1+{\ifx\relax#1\empty\else%
    \PY@tok{#1}\expandafter\PY@toks\fi}
\def\PY@do#1{\PY@bc{\PY@tc{\PY@ul{%
    \PY@it{\PY@bf{\PY@ff{#1}}}}}}}
\def\PY#1#2{\PY@reset\PY@toks#1+\relax+\PY@do{#2}}

\expandafter\def\csname PY@tok@gd\endcsname{\def\PY@tc##1{\textcolor[rgb]{0.63,0.00,0.00}{##1}}}
\expandafter\def\csname PY@tok@gu\endcsname{\let\PY@bf=\textbf\def\PY@tc##1{\textcolor[rgb]{0.50,0.00,0.50}{##1}}}
\expandafter\def\csname PY@tok@gt\endcsname{\def\PY@tc##1{\textcolor[rgb]{0.00,0.27,0.87}{##1}}}
\expandafter\def\csname PY@tok@gs\endcsname{\let\PY@bf=\textbf}
\expandafter\def\csname PY@tok@gr\endcsname{\def\PY@tc##1{\textcolor[rgb]{1.00,0.00,0.00}{##1}}}
\expandafter\def\csname PY@tok@cm\endcsname{\let\PY@it=\textit\def\PY@tc##1{\textcolor[rgb]{0.25,0.50,0.50}{##1}}}
\expandafter\def\csname PY@tok@vg\endcsname{\def\PY@tc##1{\textcolor[rgb]{0.10,0.09,0.49}{##1}}}
\expandafter\def\csname PY@tok@vi\endcsname{\def\PY@tc##1{\textcolor[rgb]{0.10,0.09,0.49}{##1}}}
\expandafter\def\csname PY@tok@vm\endcsname{\def\PY@tc##1{\textcolor[rgb]{0.10,0.09,0.49}{##1}}}
\expandafter\def\csname PY@tok@mh\endcsname{\def\PY@tc##1{\textcolor[rgb]{0.40,0.40,0.40}{##1}}}
\expandafter\def\csname PY@tok@cs\endcsname{\let\PY@it=\textit\def\PY@tc##1{\textcolor[rgb]{0.25,0.50,0.50}{##1}}}
\expandafter\def\csname PY@tok@ge\endcsname{\let\PY@it=\textit}
\expandafter\def\csname PY@tok@vc\endcsname{\def\PY@tc##1{\textcolor[rgb]{0.10,0.09,0.49}{##1}}}
\expandafter\def\csname PY@tok@il\endcsname{\def\PY@tc##1{\textcolor[rgb]{0.40,0.40,0.40}{##1}}}
\expandafter\def\csname PY@tok@go\endcsname{\def\PY@tc##1{\textcolor[rgb]{0.53,0.53,0.53}{##1}}}
\expandafter\def\csname PY@tok@cp\endcsname{\def\PY@tc##1{\textcolor[rgb]{0.74,0.48,0.00}{##1}}}
\expandafter\def\csname PY@tok@gi\endcsname{\def\PY@tc##1{\textcolor[rgb]{0.00,0.63,0.00}{##1}}}
\expandafter\def\csname PY@tok@gh\endcsname{\let\PY@bf=\textbf\def\PY@tc##1{\textcolor[rgb]{0.00,0.00,0.50}{##1}}}
\expandafter\def\csname PY@tok@ni\endcsname{\let\PY@bf=\textbf\def\PY@tc##1{\textcolor[rgb]{0.60,0.60,0.60}{##1}}}
\expandafter\def\csname PY@tok@nl\endcsname{\def\PY@tc##1{\textcolor[rgb]{0.63,0.63,0.00}{##1}}}
\expandafter\def\csname PY@tok@nn\endcsname{\let\PY@bf=\textbf\def\PY@tc##1{\textcolor[rgb]{0.00,0.00,1.00}{##1}}}
\expandafter\def\csname PY@tok@no\endcsname{\def\PY@tc##1{\textcolor[rgb]{0.53,0.00,0.00}{##1}}}
\expandafter\def\csname PY@tok@na\endcsname{\def\PY@tc##1{\textcolor[rgb]{0.49,0.56,0.16}{##1}}}
\expandafter\def\csname PY@tok@nb\endcsname{\def\PY@tc##1{\textcolor[rgb]{0.00,0.50,0.00}{##1}}}
\expandafter\def\csname PY@tok@nc\endcsname{\let\PY@bf=\textbf\def\PY@tc##1{\textcolor[rgb]{0.00,0.00,1.00}{##1}}}
\expandafter\def\csname PY@tok@nd\endcsname{\def\PY@tc##1{\textcolor[rgb]{0.67,0.13,1.00}{##1}}}
\expandafter\def\csname PY@tok@ne\endcsname{\let\PY@bf=\textbf\def\PY@tc##1{\textcolor[rgb]{0.82,0.25,0.23}{##1}}}
\expandafter\def\csname PY@tok@nf\endcsname{\def\PY@tc##1{\textcolor[rgb]{0.00,0.00,1.00}{##1}}}
\expandafter\def\csname PY@tok@si\endcsname{\let\PY@bf=\textbf\def\PY@tc##1{\textcolor[rgb]{0.73,0.40,0.53}{##1}}}
\expandafter\def\csname PY@tok@s2\endcsname{\def\PY@tc##1{\textcolor[rgb]{0.73,0.13,0.13}{##1}}}
\expandafter\def\csname PY@tok@nt\endcsname{\let\PY@bf=\textbf\def\PY@tc##1{\textcolor[rgb]{0.00,0.50,0.00}{##1}}}
\expandafter\def\csname PY@tok@nv\endcsname{\def\PY@tc##1{\textcolor[rgb]{0.10,0.09,0.49}{##1}}}
\expandafter\def\csname PY@tok@s1\endcsname{\def\PY@tc##1{\textcolor[rgb]{0.73,0.13,0.13}{##1}}}
\expandafter\def\csname PY@tok@dl\endcsname{\def\PY@tc##1{\textcolor[rgb]{0.73,0.13,0.13}{##1}}}
\expandafter\def\csname PY@tok@ch\endcsname{\let\PY@it=\textit\def\PY@tc##1{\textcolor[rgb]{0.25,0.50,0.50}{##1}}}
\expandafter\def\csname PY@tok@m\endcsname{\def\PY@tc##1{\textcolor[rgb]{0.40,0.40,0.40}{##1}}}
\expandafter\def\csname PY@tok@gp\endcsname{\let\PY@bf=\textbf\def\PY@tc##1{\textcolor[rgb]{0.00,0.00,0.50}{##1}}}
\expandafter\def\csname PY@tok@sh\endcsname{\def\PY@tc##1{\textcolor[rgb]{0.73,0.13,0.13}{##1}}}
\expandafter\def\csname PY@tok@ow\endcsname{\let\PY@bf=\textbf\def\PY@tc##1{\textcolor[rgb]{0.67,0.13,1.00}{##1}}}
\expandafter\def\csname PY@tok@sx\endcsname{\def\PY@tc##1{\textcolor[rgb]{0.00,0.50,0.00}{##1}}}
\expandafter\def\csname PY@tok@bp\endcsname{\def\PY@tc##1{\textcolor[rgb]{0.00,0.50,0.00}{##1}}}
\expandafter\def\csname PY@tok@c1\endcsname{\let\PY@it=\textit\def\PY@tc##1{\textcolor[rgb]{0.25,0.50,0.50}{##1}}}
\expandafter\def\csname PY@tok@fm\endcsname{\def\PY@tc##1{\textcolor[rgb]{0.00,0.00,1.00}{##1}}}
\expandafter\def\csname PY@tok@o\endcsname{\def\PY@tc##1{\textcolor[rgb]{0.40,0.40,0.40}{##1}}}
\expandafter\def\csname PY@tok@kc\endcsname{\let\PY@bf=\textbf\def\PY@tc##1{\textcolor[rgb]{0.00,0.50,0.00}{##1}}}
\expandafter\def\csname PY@tok@c\endcsname{\let\PY@it=\textit\def\PY@tc##1{\textcolor[rgb]{0.25,0.50,0.50}{##1}}}
\expandafter\def\csname PY@tok@mf\endcsname{\def\PY@tc##1{\textcolor[rgb]{0.40,0.40,0.40}{##1}}}
\expandafter\def\csname PY@tok@err\endcsname{\def\PY@bc##1{\setlength{\fboxsep}{0pt}\fcolorbox[rgb]{1.00,0.00,0.00}{1,1,1}{\strut ##1}}}
\expandafter\def\csname PY@tok@mb\endcsname{\def\PY@tc##1{\textcolor[rgb]{0.40,0.40,0.40}{##1}}}
\expandafter\def\csname PY@tok@ss\endcsname{\def\PY@tc##1{\textcolor[rgb]{0.10,0.09,0.49}{##1}}}
\expandafter\def\csname PY@tok@sr\endcsname{\def\PY@tc##1{\textcolor[rgb]{0.73,0.40,0.53}{##1}}}
\expandafter\def\csname PY@tok@mo\endcsname{\def\PY@tc##1{\textcolor[rgb]{0.40,0.40,0.40}{##1}}}
\expandafter\def\csname PY@tok@kd\endcsname{\let\PY@bf=\textbf\def\PY@tc##1{\textcolor[rgb]{0.00,0.50,0.00}{##1}}}
\expandafter\def\csname PY@tok@mi\endcsname{\def\PY@tc##1{\textcolor[rgb]{0.40,0.40,0.40}{##1}}}
\expandafter\def\csname PY@tok@kn\endcsname{\let\PY@bf=\textbf\def\PY@tc##1{\textcolor[rgb]{0.00,0.50,0.00}{##1}}}
\expandafter\def\csname PY@tok@cpf\endcsname{\let\PY@it=\textit\def\PY@tc##1{\textcolor[rgb]{0.25,0.50,0.50}{##1}}}
\expandafter\def\csname PY@tok@kr\endcsname{\let\PY@bf=\textbf\def\PY@tc##1{\textcolor[rgb]{0.00,0.50,0.00}{##1}}}
\expandafter\def\csname PY@tok@s\endcsname{\def\PY@tc##1{\textcolor[rgb]{0.73,0.13,0.13}{##1}}}
\expandafter\def\csname PY@tok@kp\endcsname{\def\PY@tc##1{\textcolor[rgb]{0.00,0.50,0.00}{##1}}}
\expandafter\def\csname PY@tok@w\endcsname{\def\PY@tc##1{\textcolor[rgb]{0.73,0.73,0.73}{##1}}}
\expandafter\def\csname PY@tok@kt\endcsname{\def\PY@tc##1{\textcolor[rgb]{0.69,0.00,0.25}{##1}}}
\expandafter\def\csname PY@tok@sc\endcsname{\def\PY@tc##1{\textcolor[rgb]{0.73,0.13,0.13}{##1}}}
\expandafter\def\csname PY@tok@sb\endcsname{\def\PY@tc##1{\textcolor[rgb]{0.73,0.13,0.13}{##1}}}
\expandafter\def\csname PY@tok@sa\endcsname{\def\PY@tc##1{\textcolor[rgb]{0.73,0.13,0.13}{##1}}}
\expandafter\def\csname PY@tok@k\endcsname{\let\PY@bf=\textbf\def\PY@tc##1{\textcolor[rgb]{0.00,0.50,0.00}{##1}}}
\expandafter\def\csname PY@tok@se\endcsname{\let\PY@bf=\textbf\def\PY@tc##1{\textcolor[rgb]{0.73,0.40,0.13}{##1}}}
\expandafter\def\csname PY@tok@sd\endcsname{\let\PY@it=\textit\def\PY@tc##1{\textcolor[rgb]{0.73,0.13,0.13}{##1}}}

\def\PYZus{\char`\_}

\def\PYZgt{\char`\>}
\def\PYZsh{\char`\#}

\def\PYZdq{\char`\"}

\makeatother

\newcommand{\figrule}{\hrule width \hsize height .33pt}
\newcommand{\coderule}{%
  \ifvmode\ifdim\prevdepth>-1000pt\vspace{-0.5em}\fi\else\vspace{-0.5em}\fi
  \figrule\vspace{0.2em}}

\def\Snospace~{\S{}}

\newcommand{\cc}[1]{\mbox{\smaller[0.5]\texttt{#1}}}
\newcommand{\PP}[1]{
\vspace{2px}
\noindent{\bf \IfEndWith{#1}{.}{#1}{#1.}}
}

\usepackage{pifont}
\definecolor{Gray}{gray}{0.9}

\newcommand{\boxbeg}{
\vspace{2px}
\noindent\begin{tabular}{|l|}\hline
\begin{minipage}{3.2in}
\vspace{2px}
\noindent
}

\newcommand{\boxend}{
\vspace{2px}
\end{minipage}\\ \hline
\end{tabular}
\vspace{-10pt}
}

\newcommand{\sys}{\mbox{\textsc{Bugstone-E2E}}\xspace}
\fvset{fontsize=\scriptsize,xleftmargin=14pt,numbers=left,numbersep=5pt}
\input{code/fmt}

\newcommand{\bugskills}{\mbox{bug detection skills}\xspace}
\newcommand{\phasea}{Phase~A\xspace}
\newcommand{\phaseb}{Phase~B\xspace}
\newcommand{\phaseao}{Phase~A.1\xspace}
\newcommand{\phasec}{Phase~C\xspace}

\newcommand{\phased}{Phase~D\xspace}
\newcommand{\buggy}{\cc{BUGGY}\xspace}
\newcommand{\confexp}{\cc{CONFIRMED\_EXPLOITABLE}\xspace}

\newcommand{\fp}{\cc{FALSE\_POSITIVE}\xspace}
\newcommand{\unknown}{\cc{UNKNOWN}\xspace}

\newcommand{\ra}[1]{\renewcommand{\arraystretch}{#1}}
\newcommand{\tightitems}{\begin{itemize}[leftmargin=*,noitemsep,topsep=2pt]}
\newcommand{\tightend}{\end{itemize}}

\begin{document}

% USENIX Security title block.
%
% DOUBLE-BLIND SUBMISSION: no author names, no affiliation and no e-mail
% address may appear here.  \author is deliberately empty (usenix.sty still
% typesets its 12pt italic author tabular, which collapses to nothing) and
% \date{} stops LaTeX printing today's date.  There is no ICML-style
% "Under review ... Do not distribute." notice in the USENIX style.
%
% For the camera-ready version, fill \author in as:
%   \author{
%   {\rm First Author}\\ Their Institution
%   \and
%   {\rm Second Author}\\ Their Institution
%   }
%
% ICML \icmlkeywords{} has no USENIX equivalent; the keyword list is dropped.
% (It was: vulnerability detection, CVE, program analysis, security agents,
% large language models.)

\date{}

\title{\Large \bf The History Is the Detector: Executing CVE Patch History, End-to-End}

\author{
{\rm Qiushi Wu \quad Kevin Eykholt \quad Youngja Park \quad Xiaokui Shu}\\
{\rm Dhilung Kirat \quad Douglas Lee Schales \quad Ian Molloy}\\[0.5ex]
IBM Research\\
%\texttt{\{qiushi.wu, kheykholt, email3,email4,email5,email6,email7\}@ibm.com}
}
\maketitle

\sloppy
\begin{abstract}
Public vulnerability databases collect rich information about known software flaws, including their weakness types, affected components, and related patches. Additionally, fixing commits provide the exact code changes that removed these flaws. While these records capture why the original code was unsafe, they are documented mainly for human inspection rather than automated reuse. Consequently, the same unsafe conditions may still exist elsewhere in code with no known advisory, leaving much of this detection knowledge unused.

We present \sys, a framework that automatically transforms vulnerability history into executable detection rules and validates their findings.
First, \sys mines reusable rules from verified fixing commits, capturing scan anchors, fix semantics, and CVE provenance and  organize them by CWE and language.
Second, detection follows a funnel-shaped pipeline; early stages process a large pool of candidates using lightweight analysis, while later stages apply increasingly capable and expensive models to a shrinking set of targets. 
Specifically, %\phasea 
\sys first enumerates call sites matching rule anchors using \textit{Tree-sitter}, %\phaseao 
then removes benign sites using lightweight heuristics without LLM calls. %and \phaseb 
Next, LLM-based agents inspect the remaining candidates guided by the rule.
%After that, \phasec 
Following this inspection, the system re-triages surviving candidates and builds runtime verifications, 
%while \phased 
and finally generates scope-checked patches that are validated via two-sided differential tests.
Using 19,325 high-severity CVEs from 2022 to 2026, \sys identifies 2,710 fixing commits and constructs 1,033 detection rules across 56 CWE families, packaged into 172 skills. When applied across 14 programs, it produced runtime evidence for 644 findings. These results demonstrate that CVE history can be turned into an executable workflow, transforming past vulnerabilities to reproducible detection and repair.
\end{abstract}

\section{Introduction}
\label{s:intro}

Modern vulnerability response leaves a detailed public record.  Specifically, a CVE entry describes the affected behavior and severity, a CWE label identifies the underlying weakness class, and a fixing commit usually shows the code-level changes that mitigated the risk~\cite{cvelistv5,cwe}.  For a human reviewer, these records explain what went wrong.  However, they are of little use to automated scanners because the lessons embedded in a patch do not translate into detectors capable of searching the next codebase.

\PP{The missed opportunity}
Many vulnerability patterns recur across projects and languages. For example, a command-injection fix may reveal that attacker-controlled data reaches a shell sink without proper quoting. Although API names and code structure may differ, the underlying security condition often remains the same. Prior work detects recurring vulnerabilities using known vulnerable code or patch-enhanced signatures~\cite{kim2017vuddy,xiao2020mvp}, while recent systems combine LLMs with retrieved vulnerability knowledge or program-analysis context~\cite{du2024vulrag,li2025iris,lekssays2025llmxcpg}. VulGenie further extracts API security rules from Java security patches~\cite{chen2026vulgenie}. 
However, what remains missing is a general pipeline that transforms CVE fixes across various weakness classes and languages into reusable, executable rules, applies them at scale, and validates and repairs the findings they produce.

\PP{Key challenges}
Turning these public vulnerability records into executable detectors raises three primary challenges. 
First, CVE records and patches vary widely in content, making it difficult to extract the precise components that can be reused for detection. The relevant security fixes may be buried in unrelated edits, poorly described, or tied to project-specific logic. A mining tool must therefore identify what made the original code unsafe and generalize that condition into anchors that can match the same flaw elsewhere.
Second, this knowledge must be represented as an executable rule with sufficient context to distinguish vulnerable patterns from benign cases. For instance, a simple match on \cc{system}, \cc{pickle.loads}, or \cc{memcpy} is too broad, whereas an overly specific signature risks missing the same vulnerability hidden behind a different wrapper or code structure. 
Third, these rules must scale efficiently to handle large codebases without incurring expensive analysis on every candidate. Static analyzers such as CodeQL and Semgrep scale well once suitable queries or rules exist~\cite{codeql,semgrep}, but creating and maintaining such rules still requires manual efforts.

\PP{From detection to verified remediation}
Solving the three challenges above remains insufficient for end-to-end vulnerability analysis. While a rule violation or plausible taint path identifies a potential vulnerability, it does not prove that the behavior is reachable or exploitable at runtime. A complete workflow therefore requires dynamic validation that produces concrete runtime evidence. 
Furthermore, it must support remediation by generating a feasible patch from the vulnerability evidence and program context, and then verifying that the patch removes the demonstrated behavior.

\PP{Our approach}
\sys addresses this gap through a staged pipeline spanning four core phases (\cc{Phase A--D}) that transitions seamlessly from reusable vulnerability knowledge to automated detection, runtime validation, and verified remediation. 
The system first mines known vulnerabilities from the \cc{cvelistV5} dataset, verifies their fixing commits, and synthesizes reusable rules containing \cc{scan_apis}, buggy and fixed code patterns, %a \phaseb 
a verification contract, and related source CVEs. %\cc{source_cves}. 
The rules are grouped by CWE and language into modular \bugskills.

For a target project, \phasea first indexes the codebase with Tree-sitter~\cite{treesitter} and enumerates all call sites match rule anchors without using an LLM. \phasea then removes clearly benign candidates via deterministic checks. The remaining candidates are analyzed independently in \phaseb, where an LLM-based agent uses the rule-specific verification contract to inspect taint flows, callers, guards, and sanitizers, finally returning a structured verdict accompanied by concrete code evidence. This funnel reduces the number of candidates before applying more expensive analysis and allows \phaseb to run in parallel with cheaper models.

However, static evidence alone is insufficient for full vulnerability confirmation. After \phaseb filters out most candidates, \phasec re-triages the small set of survivors and attempts to reproduce each one in an isolated, concrete execution environment. A finding gains definite runtime validation only when an expected runtime marker--such as a  sanitizer report, crash, or controlled-sink signal--is observed, and only the subset with a working live proof-of-concept (PoC) is rated as \confexp, separating non-exploitable flaws from those that could not be tested.
%This keeps not exploitable distinct from could not be tested.

Finally, \phased generates a minimal, scope-checked patch and verifies it against the runtime evidence produced in \phasec. The exploit must fail on the patched program. It then reverts the patch and confirms that the exploit succeeds again, preventing a broken build or environment from being accepted as a fix. Exploit- and test-based validation is increasingly used in recent literature to evaluate vulnerability repair and patch correctness~\cite{wang2025vulnrepaireval,bui2026vul4py,ghebremichael2026veriport}. 
However, unlike these existing settings, \sys begins with newly discovered findings that have no predefined PoC or validation environment. It must therefore construct the runtime evidence by itself and independently verify each critical artifact before accepting a patch.

\PP{Evidence}
In this work, \sys surveys 19,325 high-severity CVEs from 2022 to 2026, enriches 5,902 unique open-source CVEs, and verifies 2,710 fixing commits. From these commits, it constructs 1,033 rules across 56 CWE families, packaged into 172 production \bugskills. Across scans of 14 real-world projects, \phaseb reports 2,933 deduplicated findings. Finally, \phasec successfully obtains runtime evidence for 644 of these findings, including AddressSanitizer (ASan~\cite{serebryany2012addresssanitizer}) violations and crashes. %including 61 that satisfy the \confexp criterion. %Finally, \phased evaluates 23 findings with live proofs of concept, successfully patches all 23, and verifies each patch with a two-sided differential test. The resulting patches change a median of 8 lines in one file and cost \$0.74 per finding.

This paper makes the following contributions:
\tightitems
    %\item \textbf{Mining detection rules from CVE history.}
    \item \textbf{Vulnerability Rule Mining from CVE history:} We design a general pipeline that surveys high-severity CVEs, verifies fixing commits, extracts each patch's vulnerable sink and fix semantics, and rejects cases that do not generalize.
    %\item \textbf{Executable rules that drive cost-ordered detection.} 
    \item \textbf{Cost-ordered Detection Rule Execution:}
    Our CWE and language-specific rules carry scan anchors, buggy and fixed patterns, a verification contract, and source-CVE provenance, all packaged into modular deployable skills.  This architecture leverages deterministic indexing and lightweight filters to absorb scale, allowing LLM-based agents to apply expensive semantic judgment only where a rule points.
    %\item \textbf{One end-to-end workflow with execution as its oracle.} 
  \item \textbf{Oracle-driven End-to-End Workflow:}  Our system executes a batched re-triage and an environment-first exploit stage that separates ``not exploitable'' from ``could not be tested'', and judges patches using a two-sided differential testing under a machine-checked output contract.  We report empirical evidence for every pipeline phase and state what each result does not establish.
    %\item \textbf{Efficiency through deterministic decomposition.} 
     \item \textbf{Deterministic Decomposition for Scalability:} In \sys, anchor enumeration and the application of a frozen filter are deterministic. Thus, we can split
     a very large candidate pool into bounded, independent batches grouped by rule and source file.  Since the batches share no mutable execution state, the expensive LLM-based agentic stage can parallelize freely--even though candidates within a single batch are not statistically independent--resulting in high scalability.
\tightend

\section{Background and Problem}
\label{s:study}

%%--------------------------------------------------------------------
\subsection{Motivation}
\label{ss:motivation}

\PP{Vulnerability patterns recur across projects}
Known vulnerability patterns often reappear in other code locations or projects. Early systems, such as ReDeBug and VUDDY, demonstrated this problem by finding unpatched clones of known vulnerable codebases at large scale~\cite{jang2012redebug,kim2017vuddy}. Later work generalized beyond exact clones by deriving vulnerability and patch signatures that identify recurring vulnerabilities despite code changes~\cite{xiao2020mvp,woo2023v1scan,huang2024vmud}. More recently, Wu et al.~\cite{wu2025onebug} showed that identical security-relevant usage patterns often recur many times within a single software project. Starting with 135 rules, their study flagged 22,568 potential violations in the Linux kernel, where a subsequent manual review of 400 cases confirmed 246 true vulnerabilities.

\PP{Cross-language recurrence}
Prior evidence has focused largely on code reuse and recurring patterns within a particular language or software ecosystem. To examine whether the same property holds more broadly, we study public CVE fixing commits across weakness classes and programming languages. \autoref{t:top-rules} reports representative high-recurrence patterns for different languages. We observe that many patterns are supported by multiple independent CVEs, rather than a single project-specific fix. We also find that the same weakness class can recur across languages through similar security-relevant operations. For example, cross-site scripting (CWE-79) appears across PHP, JavaScript, and TypeScript, while command injection (CWE-78) affects Python and Ruby.

CWE-78-Py-R1, our running example, is supported by thirteen CVEs from different Python projects. The common pattern is that externally derived data reaches a shell API without sufficient quoting or sanitization. In one instance from PaddleOCR, a path selected through \cc{QFileDialog} is used to derive an output directory and then concatenated directly into an \cc{os.system} command. Although the surrounding application logic differs across projects, the security-relevant condition is the same. This recurrence motivates extracting such conditions from CVE fixes and reusing them to guide detection in previously unseen code.

\begin{table}[t]
\centering
\footnotesize
\ra{1.05}
\caption{Representative high-recurrence patterns from our CVE background
  study, spanning ten of the twelve languages \sys covers.  \#CVEs counts
  distinct CVEs whose patches support the pattern, and the same weakness class
  recurs across languages.}
\label{t:top-rules}
\begin{tabular}{@{}llrl@{}}
\toprule
\textbf{CWE} & \textbf{Lang.} &
  \textbf{\#CVEs} & \textbf{Repr.\ API} \\
\midrule
CWE-79  & PHP    & 27 & \cc{echo} \\
CWE-502 & PHP    & 16 & \cc{unserialize} \\
CWE-78  & Python & 13 & \cc{os.system} \\
CWE-79  & JS     &  9 & \cc{innerHTML} \\
CWE-639 & Go     &  7 & \cc{Decode} \\
CWE-79  & TypeScript     &  7 & \cc{src} \\
CWE-22  & Java   &  5 & \cc{getNextEntry} \\
CWE-122 & C      &  4 & \cc{memcpy} \\
CWE-122 & C++    &  4 & \cc{memcpy} \\
CWE-22  & C\#    &  4 & \cc{Path.Combine} \\
CWE-78  & Ruby   &  2 & \cc{system} \\
\bottomrule
\end{tabular}
\end{table}

\iffalse
\begin{figure}[t]
\centering
\small
\coderule
\input{code/injection.py.tex}
\coderule
\caption{Representative command-injection pattern.  A path from a UI
  dialog reaches \cc{os.system} via string concatenation, and the
  CVE-derived rule captures the sink and the missing-sanitization
  condition rather than project specifics.}
\label{fig:code-example}
\end{figure}

\figureautorefname
\fi

%%--------------------------------------------------------------------
\subsection{Challenges in Precise and Scalable Vulnerability Detection}
\label{ss:challenges}

\PP{Traditional static analysis}
Static analyzers such as CodeQL~\cite{codeql} and Semgrep~\cite{semgrep} can efficiently scan large codebases once suitable queries or rules are available. However, encoding a new vulnerability pattern still requires analyst effort, which limits how quickly these tools can absorb the broad and evolving set of patterns recorded in CVE history. Their findings also often require additional context to distinguish true vulnerabilities from benign uses, and false positives remain a major barrier to developer adoption~\cite{johnson2013dont}. Recent work therefore combines static analysis with semantic reasoning to improve the triage of candidate findings~\cite{li2025iris}. Static analysis provides scalable candidate discovery, but extending its coverage and resolving ambiguous findings still require substantial manual or semantic analysis.

\PP{Dynamic analysis and fuzzing}
Dynamic techniques provide stronger evidence because they observe concrete program behavior. Tools such as ASan~\cite{serebryany2012addresssanitizer} detect memory errors during execution, while fuzzers such as AFL++~\cite{fioraldi2020aflplusplus} explore program behavior through generated inputs. Their effectiveness, however, depends on both reaching the vulnerable behavior and having an oracle that recognizes it. Memory errors often produce crashes or sanitizer violations, but many other vulnerabilities do not. For command injection, path traversal, deserialization, and similar weaknesses, successful execution may appear normal unless the analysis encodes the relevant security condition. Dynamic validation is therefore valuable for confirming vulnerabilities, but it is difficult to use as the primary mechanism for broad vulnerability discovery.

\PP{LLM and agent-based detection}
Recent work uses LLMs to provide semantic context that is difficult to encode in fixed rules. VulRAG~\cite{du2024vulrag} retrieves vulnerability knowledge to guide analysis, IRIS~\cite{li2025iris} and LLift~\cite{li2024llift} combine LLM reasoning with static-analysis results, and LLMxCPG~\cite{lekssays2025llmxcpg} uses code property graphs to provide structured program context. More recent systems employ agentic workflows that can inspect code, reason about attack surfaces, and validate candidate vulnerabilities~\cite{openaicodexsec,anthropicdefcode,visavvah,bigsleep2024naptime,kim2025atlantis,zhang2026aixcc}.
These systems improve semantic reasoning but introduce a different scaling problem. Repository-wide exploration gives the agent a large search space and little guidance about where expensive reasoning is most useful. Deeper analysis of one region consumes budget that could otherwise cover additional code, and the result depends strongly on the capability of the underlying model. Moreover, an individual scan usually produces findings rather than reusable detection knowledge, so similar reasoning may need to be repeated on the next target. This motivates a different design: use persistent vulnerability knowledge to identify where to look, apply inexpensive analysis broadly, and reserve expensive semantic and runtime reasoning for a progressively smaller set of candidates.

%%--------------------------------------------------------------------
\subsection{Research Scope}
\label{ss:problem}

%\PP{Terminology}
%We use two terms in a specific sense throughout this paper.
%An \emph{agent} is an automated tool with tool-calling that interacts
%with an LLM, for example Claude Code, Codex, or the Pi agent.
%A \emph{harness} is the combination of an agent, an LLM, the code that
%implements the system workflow, and the vulnerability-detection skills
%and scripts.
%Under these definitions \sys and its agentic baselines are all harnesses.

%\PP{Scope}
\sys targets vulnerabilities that follow recurring API-misuse patterns, where the security-relevant behavior is visible at or near a call site. Its scope does not include one-off design flaws, configuration errors, or vulnerabilities that require whole-program symbolic reasoning to identify. \sys produces candidate findings with structured static and runtime evidence, while human-in-the-loop disclosure decisions and patch acceptance remain separate processes.

\section{System Overview}
\label{s:overview}

\begin{figure*}[t]
\centering
\resizebox{\textwidth}{!}{%
\begin{tikzpicture}[x=1cm, y=1cm]

%% ---- Card style ----
\tikzset{
  card/.style={
    rectangle, rounded corners=4pt,
    minimum width=2.1cm, minimum height=2.0cm,
    fill=white, draw=gray!35, line width=0.5pt,
    text width=1.9cm, align=center, font=\footnotesize
  }
}

%% ===================================================================
%% REGIONS
%% ===================================================================
\begin{scope}[on background layer]
  %% Part 1 outer region and two internal stages
  \fill[gray!4, rounded corners=5pt]
    (0.00, 6.95) rectangle (18.00, 10.25);
  \fill[blue!8, rounded corners=5pt]
    (0.10, 7.05) rectangle (12.45, 9.60);
  \fill[teal!10, rounded corners=5pt]
    (12.55, 7.05) rectangle (17.90, 9.60);

  %% Host and guest execution regions
  \fill[gray!6, rounded corners=5pt]
    (0.00, 1.90) rectangle (14.85, 5.95);
  \fill[orange!10, rounded corners=4pt]
    (0.15, 2.05) rectangle (7.20, 5.35);
  \fill[red!8, rounded corners=4pt]
    (7.30, 2.05) rectangle (11.85, 5.35);
  \fill[red!8, rounded corners=4pt]
    (12.60, 2.05) rectangle (14.70, 5.35);
\end{scope}

\draw[gray!25, rounded corners=5pt, line width=0.8pt]
  (0.00, 6.95) rectangle (18.00, 10.25);
\draw[gray!45, rounded corners=5pt, line width=1.0pt]
  (0.00, 1.90) rectangle (14.85, 5.95);
\draw[orange!50, rounded corners=4pt, line width=0.8pt]
  (0.15, 2.05) rectangle (7.20, 5.35);
\draw[red!45, rounded corners=4pt, line width=0.8pt]
  (7.30, 2.05) rectangle (11.85, 5.35);
\draw[red!45, dashed, rounded corners=4pt, line width=0.8pt]
  (12.60, 2.05) rectangle (14.70, 5.35);

%% ===================================================================
%% REGION LABELS
%% ===================================================================
\node[font=\small\bfseries, text=blue!65]
  at (9.00, 10.00) {Part 1: CVE Intelligence and Skill Development};

\node[font=\scriptsize\bfseries, text=blue!60]
  at (6.20, 9.42) {CVE Mining};
\node[font=\scriptsize\bfseries, text=teal!65]
  at (15.25, 9.42) {Rule \& Skill Synthesis};

\node[font=\scriptsize\bfseries, text=gray!55, anchor=north west]
  at (0.12, 5.88) {Host Agent};

\node[font=\small\bfseries, text=orange!65]
  at (3.68, 5.10) {Part 2: Rule-Driven Detection};
\node[font=\small\bfseries, text=red!55]
  at (9.58, 5.10) {Part 3: Validation \& Remediation};

\node[font=\scriptsize\bfseries, text=orange!65, anchor=south west]
  at (0.32, 2.12) {Guest VM};

%% ===================================================================
%% PART 1 NODES
%% ===================================================================
\node[card] (step1) at (1.2, 8.20)
  {\rule{0pt}{10pt}\\
   \textbf{Survey CVEs}\\
   \textcolor{gray!60}{\scriptsize 19,325 high-sev.}};
\node[fill=blue!60, anchor=north,
      minimum width=2.1cm, minimum height=8pt] at (step1.north) {};

\node[card] (step2) at (3.7, 8.20)
  {\rule{0pt}{10pt}\\
   \textbf{Identify +}\\
   \textbf{Enrich OSS}\\
   \textcolor{gray!60}{\scriptsize public repositories}};
\node[fill=blue!60, anchor=north,
      minimum width=2.1cm, minimum height=8pt] at (step2.north) {};

\node[card] (step3) at (6.2, 8.20)
  {\rule{0pt}{10pt}\\
   \textbf{Clone +}\\
   \textbf{Date Filter}\\
   \textcolor{gray!60}{\scriptsize commit window}};
\node[fill=blue!60, anchor=north,
      minimum width=2.1cm, minimum height=8pt] at (step3.north) {};

\node[card] (step4) at (8.7, 8.20)
  {\rule{0pt}{10pt}\\
   \textbf{Light LLM}\\
   \textbf{Screen}\\
   \textcolor{gray!60}{\scriptsize pre-filter patches}};
\node[fill=blue!60, anchor=north,
      minimum width=2.1cm, minimum height=8pt] at (step4.north) {};

\node[card] (step5) at (11.2, 8.20)
  {\rule{0pt}{10pt}\\
   \textbf{Verify Fix}\\
   \textbf{Commit}\\
   \textcolor{gray!60}{\scriptsize 2,710 confirmed}};
\node[fill=blue!60, anchor=north,
      minimum width=2.1cm, minimum height=8pt] at (step5.north) {};

\node[card] (step6) at (13.7, 8.20)
  {\rule{0pt}{10pt}\\
   \textbf{Rule}\\
   \textbf{Synthesis}\\
   \textcolor{gray!60}{\scriptsize extract + generalize}};
\node[fill=teal!65, anchor=north,
      minimum width=2.1cm, minimum height=8pt] at (step6.north) {};

\node[card] (step7) at (16.2, 8.20)
  {\rule{0pt}{10pt}\\
   \textbf{Skill}\\
   \textbf{Synthesis}\\
   \textcolor{gray!60}{\scriptsize 1,033 rules\\172 skills}};
\node[fill=teal!65, anchor=north,
      minimum width=2.1cm, minimum height=8pt] at (step7.north) {};

%% ---- Part 1 arrows ----
\draw[-Stealth, blue!55, thick] (step1.east) -- (step2.west);
\draw[-Stealth, blue!55, thick] (step2.east) -- (step3.west);
\draw[-Stealth, blue!55, thick] (step3.east) -- (step4.west);
\draw[-Stealth, blue!55, thick] (step4.east) -- (step5.west);
\draw[-Stealth, blue!55, thick] (step5.east) -- (step6.west);
\draw[-Stealth, teal!60, thick] (step6.east) -- (step7.west);

%% ===================================================================
%% PART 2 AND PART 3 NODES
%% ===================================================================
\node[card] (phaseA) at (1.15, 3.80)
  {\rule{0pt}{10pt}\\
   \textbf{tree-sitter}\\
   \textbf{Index}\\
   \textcolor{gray!60}{\scriptsize \phasea}};
\node[fill=orange!60, anchor=north,
      minimum width=2.1cm, minimum height=8pt] at (phaseA.north) {};

\node[card] (phaseAhalf) at (3.55, 3.80)
  {\rule{0pt}{10pt}\\
   \textbf{Filter.py}\\
   \textcolor{gray!60}{\scriptsize \phaseao}};
\node[fill=orange!60, anchor=north,
      minimum width=2.1cm, minimum height=8pt] at (phaseAhalf.north) {};

\node[card] (phaseB) at (5.95, 3.80)
  {\rule{0pt}{10pt}\\
   \textbf{LLM Agent}\\
   \textbf{Verify}\\
   \textcolor{gray!60}{\scriptsize \phaseb}};
\node[fill=orange!60, anchor=north,
      minimum width=2.1cm, minimum height=8pt] at (phaseB.north) {};

\node[card] (phaseC) at (8.35, 3.80)
  {\rule{0pt}{10pt}\\
   \textbf{Triage +}\\
   \textbf{Exploit}\\
   \textcolor{gray!60}{\scriptsize \phasec}};
\node[fill=red!55, anchor=north,
      minimum width=2.1cm, minimum height=8pt] at (phaseC.north) {};

\node[card] (phaseD) at (10.75, 3.80)
  {\rule{0pt}{10pt}\\
   \textbf{Patch +}\\
   \textbf{Verify}\\
   \textcolor{gray!60}{\scriptsize \phased}};
\node[fill=violet!60, anchor=north,
      minimum width=2.1cm, minimum height=8pt] at (phaseD.north) {};

\node[card] (report) at (13.65, 3.80)
  {\rule{0pt}{10pt}\\
   \textbf{Report +}\\
   \textbf{Package}\\
   \textcolor{gray!60}{\scriptsize JSON + HTML}};
\node[fill=gray!55, anchor=north,
      minimum width=2.1cm, minimum height=8pt] at (report.north) {};

%% ---- Forward arrows ----
\draw[-Stealth, orange!65, thick] (phaseA.east)     -- (phaseAhalf.west);
\draw[-Stealth, orange!65, thick] (phaseAhalf.east) -- (phaseB.west);
\draw[-Stealth, red!55, thick]    (phaseB.east)     -- (phaseC.west);
\draw[-Stealth, violet!60, thick] (phaseC.east)     -- (phaseD.west);
\draw[-Stealth, gray!55, thick]   (phaseD.east)     -- (report.west);

%% ===================================================================
%% FEEDBACK LOOPS
%% ===================================================================
\draw[-Stealth, dashed, orange!75, semithick]
  (phaseB.south) -- ++(0, -0.35)
  -- node[midway, below, font=\scriptsize, text=orange!75]
       {optimize filter (1--2 rounds)}
     (3.55, 2.45)
  -- (phaseAhalf.south);

\draw[-Stealth, dashed, violet!70, semithick]
  (phaseD.south) -- ++(0, -0.35)
  -- node[midway, below, font=\scriptsize, text=violet!70]
       {re-run PoC}
     (8.35, 2.45)
  -- (phaseC.south);

%% ===================================================================
%% CONNECTOR: SKILLS -> TARGET ANALYSIS
%% ===================================================================
\draw[-Stealth, teal!65, thick]
  (step7.south)
  -- (16.2, 6.45)
  -- node[midway, above, font=\scriptsize, text=teal!70]
       {skills + rules}
     (1.15, 6.45)
  -- (phaseA.north);

\end{tikzpicture}%
}
\caption{Overview of \sys.}
\label{fig:overview}
\end{figure*}

\autoref{fig:overview} summarizes \sys in three parts.
Part~1 converts public vulnerability history into reusable detection skills.
Part~2 applies those skills to a target project and identifies candidates that remain plausible after static analysis.
Part~3 raises the evidence requirement through runtime validation and verified remediation.
The pipeline therefore moves from historical vulnerability knowledge to increasingly stronger evidence while applying expensive analysis to progressively fewer candidates.

\PP{Part 1: From CVE History to Detection Skills}
\sys treats CVE fixing commits as rich sources of reusable detection knowledge rather than project-specific repairs. It surveys 19,325 high-severity CVEs and verifies 2,710 fixing commits from open-source projects. From these commits, \sys extracts reusable scan anchors, buggy and fixed patterns, verification contracts, and CVE provenance, while rejecting patterns that fail to generalize. It then consolidates related CWE categories to group rules by CWE family and language. The resulting knowledge base contains 1,033 production rules across 56 CWE families, packaged into 172 deployable \bugskills.

\PP{Part 2: Rule-Driven Detection}
Given a target repository, \phasea enumerates rule-anchored call sites, and \phaseao removes false positives using lightweight filtering, leaving a reduced candidate set for semantic verification. However, this stage can still produce hundreds to thousands of candidates for medium-to-large projects. \phaseb then decomposes this workload into independent tasks, assigning one agent to each candidate together with the corresponding rule. Instead of exploring or reasoning about the repository as a whole, each agent only needs to determine whether that specific candidate violates the rule and constitutes the reported bug pattern. This narrow task scope substantially lowers the required agent capability, enabling effective verification using smaller, more cost-efficient models. Because these candidates are independent, \phaseb can also process them with high parallelism, scaling semantic analysis to large candidate sets without requiring a shared state.

\PP{Part 3: Validation and Remediation}
The output of Part~2 is still a static claim, so Part~3 applies stronger evidence requirements only to the surviving candidates from \phaseb.
\phasec first re-triages these findings in batches and then attempts to construct a concrete execution environment for runtime validation.
It records runtime evidence such as sink reachability, sanitizer violations, or crashes, while only findings that satisfy the stronger exploit-confirmation criterion are marked \confexp.
Failure to construct or exercise an environment remains distinct from evidence that a finding is not exploitable.
This ordering keeps expensive runtime analysis focused on the small set of candidates that survive earlier stages.

For findings with sufficient runtime evidence, \phased generates a minimal, scope-checked patch using the vulnerability evidence and program context.
It then applies a two-sided differential test: the proof of concept must fail on the patched tree and succeed again after the patch is reverted.
This prevents a broken build or invalid environment from being accepted as a successful repair.
Finally, \sys follows the project documentation to produce a vulnerability report in the maintainer's required format, together with supporting evidence and a proposed patch.

%Overall, the three stages establish progressively stronger claims:
%\phaseb provides structured static evidence that a rule is violated,
%\phasec establishes runtime evidence and exploitability,
%and \phased establishes a verified remediation.

\section{Design}
\label{s:design}

\sys is designed around three problems that arise when turning vulnerability
history into an end-to-end detection workflow.
First, public CVE records and fixing commits must be converted into reusable
detection knowledge without overfitting to one patch or learning the fix rather
than the vulnerability.
Second, that knowledge must guide semantic analysis at repository scale without
requiring expensive model reasoning over the entire codebase.
Third, findings and generated patches must be supported by evidence that does
not depend on the model's own judgment.
These problems motivate the three research challenges below.
The following subsections then describe how
\sys addresses them.

\subsection{Research Challenges}
\label{ss:design-challenges}

\PP{RC1 (Turning vulnerability history into reusable detection
knowledge).}
A CVE record is not directly usable as a detection specification.
Vulnerability descriptions are often brief, fixing commits may not be
linked explicitly, and patches can mix the security fix with unrelated changes.
Even when the correct fixing commit is known, the changed line does not
necessarily identify the underlying vulnerability condition. 
For example, a patch may place a
check, sanitizer, or guard away from the actual vulnerable operation.
Directly translating such diffs into rules can therefore produce detectors for
the fix rather than for the vulnerable code.
Additionally, generating one rule per CVE results in highly redundant, project-specific rules, whereas merging them aggressively can combine distinct root causes.
The challenge is to isolate and validate the relevant fix, identify the
security-relevant condition, and generalize it only when supported by multiple instances.

\PP{RC2 (Scaling semantic analysis through task decomposition).}
Applying an extensive rule set to a large codebase can produce an enormous amount of semantic analysis.
While rules provide explicit candidate locations, determining whether a candidate is truly vulnerable may require inspecting surrounding data flows, guards, sanitization,
and project-specific conventions.
Simply assigning one agent to each rule does not solve this problem. 
A single rule can match hundreds or thousands of sites, overwhelming the agent with a large
context and a long sequence of rule-specific judgments.
Such tasks place substantial demands on both model capability and context
handling.
Therefore, the challenge is to minimize the number of candidates that
reach semantic analysis and to break down the remaining work into small,
independent reasoning tasks.

\PP{RC3 (Establishing evidence beyond model judgment).}
A model's conclusion that a candidate violates a rule is merely a static claim, not runtime evidence of an acutal vulnerability.
Likewise, a patch generated by the same model cannot be considered correct
simply because the model reports that it works.
Both stages require evidence that can be checked independently of the model.
For vulnerability validation, the execution must distinguish observed runtime
evidence, stronger exploit confirmation, unsuccessful tests, and cases where
the target cannot be exercised.
For remediation, verification must demonstrate that the triggering input no longer succeeds after the patch, but the same input succeeds when the patch is reverted.
The challenge is therefore to raise the burden of proof from structured static
evidence, to runtime evidence, and finally to verified remediation while
applying the more expensive checks only to the progressively smaller set of
surviving findings, none of which is taken on the reporting agent's word.

\subsection{CVE-to-Skill Pipeline}
\label{ss:cve-rule}

\begin{table}[h]
\centering
\scriptsize
\ra{1.08}
\setlength{\tabcolsep}{4pt}
\begin{tabular}{
p{0.30\columnwidth}
p{0.40\columnwidth}
p{0.20\columnwidth}}
\toprule
\textbf{Stage} & \textbf{Artifact} & \textbf{Count} \\
\midrule

CVE survey &
High-severity CVEs &
19,325 \\
\midrule

OSS enrichment &
Open-source CVEs &
5,902 \\
\midrule

Fix recovery &
Verified fixing commits &
2,710 \\
\midrule

Case validation &
Validated CVE cases &
2,662 \\
\midrule

Rule synthesis &
Pre-consolidation rules / CWEs &
1,757 / 266 \\
\midrule

Family consolidation &
Production rules / CWE families &
1,033 / 56 \\
\midrule

Skill synthesis &
Deployable \bugskills &
172 \\
\bottomrule

\end{tabular}

\caption{CVE-to-skill pipeline and the number of artifacts retained at each stage.}
\label{t:cve-pipeline}
\end{table}

To address RC1, \sys turns incomplete CVE records into deployable detection
skills in three steps: it enriches CVE metadata and recovers verified fixing
commits, extracts reusable vulnerability conditions from those fixes, and
consolidates related rules into CWE-family--language skills.
\autoref{t:cve-pipeline} summarizes this progression.

\PP{CVE enrichment and fix recovery.}
CVE metadata is often insufficient to identify either the affected code or its
fix.
A record may contain only a short vulnerability description, omit the affected
repository or fixing commit, or reference an issue, advisory, or third-party
page rather than the relevant code change.
CWE metadata can also be missing, ambiguous, or contain several weakness
categories in one vulnerability.

\sys therefore enriches each record before rule extraction.
Starting from 19,325 high-severity CVEs published between 2022 and 2026, it
combines structured CVE fields, natural-language descriptions, referenced
pages, and public repository information to resolve the affected open-source
project and primary weakness category.
When a reliable fixing-commit reference is available, \sys verifies the cited
commit directly.
Otherwise, it uses the vulnerability reporting and publication window to guide
an agentic search over repository history.
Candidate commits are checked against the CVE description, patch, and
surrounding code to determine whether they actually fix the reported
vulnerability.
This process yields 2,710 verified fixing commits, which form the evidence base
for rule generation.

\PP{Patch-guided rule extraction.}
A verified fixing commit is still not a detection rule.
A patch may mix the security fix with unrelated changes, and the code added by
the fix may be a guard or sanitizer rather than the operation that was
originally unsafe.
The goal is therefore to recover the underlying vulnerability condition rather
than reproduce the textual shape of the patch.

For each verified fix, \sys jointly analyzes the CVE context, the vulnerable
and fixed code, and the surrounding functions.
From this evidence, it identifies the vulnerable operation, candidate scan
anchors, the condition that makes the operation unsafe, and the change that
removes that condition.
The resulting rule describes what should be detected in vulnerable code rather
than what was introduced by the fix.
A validation step checks this direction explicitly: scan anchors must refer to
operations present before the fix and must not point to a sanitizer, guard, or
replacement API added by the patch.
Rules that cannot be supported by the available evidence or generalized beyond
the individual patch are not promoted to the production rule set.
Each accepted rule retains its source CVEs so that later findings remain
traceable to the vulnerability evidence behind the rule.

This process produces 1,757 pre-consolidation rules covering 266 CWE
categories.
These numbers describe the mined rule set before related weakness categories
and overlapping rules are consolidated.

\PP{Rule consolidation and skill synthesis.}
CWE categories do not map one-to-one to distinct detection problems.
Related CWEs often describe different manifestations of the same underlying
condition and consequently share scan anchors and semantic checks.
Keeping them independent would repeatedly enumerate and analyze the same code
sites.

\sys therefore groups closely related CWE categories into broader
\emph{CWE families} and consolidates overlapping rules within each family.
The resulting production knowledge base contains 1,033 rules across 56 CWE
families, reduced from 1,757 rules across 266 CWE categories.
This consolidation removes redundant detection work while preserving rules for
distinct vulnerability conditions.

The consolidated rules are organized by CWE family and target language into
172 deployable \bugskills.
A skill is not a single rule; it is an execution template for one
CWE-family--language combination and may contain multiple rules covering
different APIs or variants of the same vulnerability family.
Its rule file %\cc{rules.json} 
stores rule-specific scan anchors and vulnerability
conditions, while a skill file (\cc{SKILL.md}) defines the semantic checks and evidence
expected from \phaseb.
Each skill also contains a filtering script %\cc{filter.py} 
for inexpensive deterministic
filtering.
This filter may initially be empty and, when candidate volume justifies it, can
be refined from \phaseb feedback during large-scale scans as described in
\autoref{ss:scanner}.
The CVE-derived rules themselves remain unchanged.

For example, the Python command-injection family is supported by multiple CVEs, where externally derived data reaches a shell-execution API without
sufficient quoting or sanitization.
Although the individual patches differ, they support the same reusable
condition.
The corresponding skill therefore anchors on the relevant shell APIs and asks
downstream analysis whether attacker-controlled data can reach them without an
adequate guard.

\subsection{Layered Target Scanner}
\label{ss:scanner}

Once \bugskills are constructed, the remaining challenge is to apply them to a
large repository without turning every rule match into an expensive semantic or
runtime analysis.
\sys therefore uses a layered scanner that raises both cost and burden of proof
as the candidate set shrinks.
\phasea performs broad build-free enumeration, \phaseao reduces redundant or
repeatedly benign candidates, \phaseb performs bounded semantic verification,
\phasec establishes runtime evidence and exploitability, and \phased verifies
remediation.

\PP{Build-free enumeration and conservative self-improvement (RC2).}
%\phasea provides broad coverage without requiring the target to build.
In \phasea, \sys parses source files with Tree-sitter~\cite{treesitter} and constructs a
shared index of call sites that is reused across all applicable skills.
Each rule contributes explicit scan anchors, and \phasea enumerates matching
call sites without invoking an LLM.
For a small target, or for a rule that matches only a modest number of sites,
these candidates can proceed directly to semantic verification.

However, large repositories create a different problem.
Common APIs such as memory allocation, deallocation, or string operations can
appear thousands of times, making semantic inspection of every matching site
unnecessarily expensive.
When this occurs, \sys can enable an iterative filter-synthesis loop (\phaseao).
A new bug detection skill may begin with an empty filter.
\phaseb then examines a bounded sample of candidates and identifies false
positives.
From these results, an agent summarizes recurring benign patterns that can be
expressed using inexpensive local syntax or AST checks and updates the filter
conservatively.
The filter is then applied to subsequent candidates, whose \phaseb results can
in turn refine it further.
Thus, semantic judgments from early candidates are distilled into cheap checks
that reduce the workload for later candidates.

The filter is deliberately asymmetric relative to the CVE-derived rule.
It only removes patterns judged safe by deterministic checks and does not
modify the rule's vulnerability condition or scan anchors.
If a filter cannot determine that a candidate is benign, the candidate is
retained.
The refined filter is stored with the skill and can be reused in
later scans.

\phaseao also removes duplicate work that arises only after rules are applied
to a target.
When overlapping rules within the same CWE family identify the same source
location, the scanner keeps a representative candidate rather than sending
multiple equivalent tasks to \phaseb.
Note that \phaseao reduces redundant work in a particular target, while rule consolidation in \autoref{ss:cve-rule} reduces redundancy in the
knowledge base.

\PP{Bounded semantic verification (RC2).}
The remaining candidates require semantic reasoning because safety may depend
on data flow, surrounding guards, sanitization, callers, or project-specific
conventions.
Assigning all candidates for one skill to a single agent would still create a
large reasoning task: one rule may match hundreds or thousands of locations,
forcing the agent to process extensive context and make many independent
judgments in one session.

\sys instead decomposes verification into small, bounded tasks.
Candidates are grouped only when they match the same rule in the same source
file, allowing them to reuse rule instructions and source context without
turning verification back into a repository-scale task.
The maximum number of candidates assigned to one worker is configurable and can
be adjusted to the capability of the \phaseb model.
Each worker receives the corresponding skill and candidate locations and may
inspect nearby code, callers, data flow, guards, and sanitizers as needed.
It returns a verdict together with structured rule-specific evidence, such as
the matched sink, relevant data flow, and guard or sanitizer analysis.
Because these tasks are small and independent, they can be processed in
parallel while placing substantially lower contextual and reasoning demands on
each model invocation.

\PP{Runtime validation and exploit confirmation (RC3).}
A \phaseb finding provides structured static evidence, but it does not establish
that the reported condition can be exercised at runtime.
\phasec therefore moves each surviving finding into an isolated dynamic
validation workflow.

The first step constructs a reusable execution environment for the target.
A LLM-based agent runs inside a dedicated VM or container, installs the
required dependencies, builds the project, and enables sanitizers when
applicable.
Because this setup is shared by all later validation tasks for the same target,
\sys performs it once and records a compact environment report describing how
the target was built and exercised.

Each \phaseb finding is then validated independently by a \phasec agent using
this prepared environment and its build report.
The agent focuses on one candidate at a time, constructs an input or execution
path that reaches the reported operation, and looks for a vulnerability-specific
runtime oracle, such as an ASan violation, crash, controlled sink reachability,
or other observable security-relevant behavior.
Dynamic validation is executed sequentially rather than concurrently to avoid
interference between concurrent tests.
After each candidate, \sys saves the resulting evidence and restores the
environment to its pre-test state before validating the next finding.
Runtime evidence is recorded separately from stronger exploit confirmation, because observing the reported behavior does not, by itself, establish a working exploit.
Thus, \phasec distinguishes among four categories: findings with runtime evidence,
exploit-confirmed findings, findings that could not be proved exploitable, and
cases that could not be tested due to an insufficient execution environment.

\PP{Differentially verified remediation (RC3).}
For findings with a working proof of concept, \phased attempts to produce a
minimal, scope-checked patch using the vulnerability report and runtime
evidence.
Patch generation alone is insufficient because the model that writes a patch
cannot certify its own fix.
A one-sided test is also weak, because the proof of concept may stop working because
the patch broke the build or disabled the relevant execution path rather than
because it removed the vulnerability.

\sys therefore uses a two-sided differential test.
It first applies the patch and reruns the same proof of concept, which must no
longer reproduce the vulnerability.
It then reverts the patch and runs the proof of concept again, which must
reproduce the original behavior.
Only patches that satisfy both directions are marked as verified.
The final output packages the maintainer-format vulnerability report together
with the proof of concept, runtime evidence, proposed patch, and verification
artifacts.
This completes the progression from structured static evidence, to runtime
evidence and exploit confirmation, and finally to verified remediation.

\section{Evaluation}
\label{s:eval}

We evaluate \sys along the main design questions introduced in
\autoref{s:design}.
We first examine how effectively the CVE-to-skill pipeline converts public
vulnerability history into deployable detection knowledge.
We then evaluate the target scanner through controlled comparisons,
reproducibility experiments, and cost measurements.
Real-world findings and runtime validation results are reported separately in
\autoref{s:realworld}.

\subsection{Experimental Setup}
\label{ss:setup}

\PP{Hardware.}
All experiments run on a dual-socket server with two Intel Xeon Gold 6336Y
processors, 96 hardware threads, and 2 TB of RAM.
\phasea uses up to 96 threads for deterministic enumeration, while model-based
stages are bounded by API rate limits and the concurrency configured for each
experiment.

\PP{Models and agent harnesses.}
Unless otherwise stated, \phaseb uses \cc{gpt-5-mini}.
Controlled experiments additionally evaluate GPT-5.5, GPT-5.6-sol, Claude
Sonnet~4.6, Kimi-K2.5, and other models where noted.
\sys supports multiple coding-agent harnesses, including Claude Code, Codex,
and Pi.
For the cross-harness study in \autoref{ss:harness}, we compare against three
recent agentic security harnesses: the OpenAI/Codex security-scan
harness~\cite{openaicodexsec}, Anthropic's open-source \emph{defending code}
reference harness~\cite{anthropicdefcode}, and the Visa Vulnerability Agentic
Harness~\cite{visavvah}.

\PP{Evaluation targets.}
We use different targets for different controlled experiments.
wolfSSL is rolled back to a pre-fix revision for the cross-harness comparison
in \autoref{ss:harness}.
Pillow is similarly rolled back and used for the reproducibility and model
studies in \autoref{ss:repro}.
The larger deployment corpus used to measure end-to-end scanning is described
separately in \autoref{s:realworld}.
This separation prevents findings accumulated across repeated production scans
from being mixed with controlled experimental results.

\subsection{CVE Knowledge Construction}
\label{ss:cvedata}

We first evaluate the CVE-to-skill pipeline of
\autoref{ss:cve-rule}.
Starting from 19,325 high-severity CVEs published between 2022 and 2026, the
pipeline identifies 5,902 records associated with public open-source
repositories and verifies 2,710 fixing commits.
Patch-guided synthesis produces 1,757 pre-consolidation rules spanning 266 CWE
categories.
Family-level consolidation reduces this knowledge base to 1,033 production
rules across 56 CWE families, organized into 172 language-specific
\bugskills.
\autoref{t:cve-pipeline} summarizes the full progression.

\PP{Recovering fixes beyond structured CVE metadata.}
Of the 19,325 high-severity CVEs surveyed, \sys enriches 6,448 records using
structured metadata, natural-language descriptions, and external references.
This process resolves 5,902 records to open-source projects with identifiable
GitHub repositories.
Among these 5,902 records, 2,439 contain a direct commit URL, from which
\sys verifies 1,944 fixing commits.
For records with a resolved repository but no usable commit pointer, \sys
searches repository history using the CVE context and vulnerability timeline,
recovering 766 additional fixing commits.
Together, the direct-reference and search paths yield 2,710 verified fixing
commits.
Thus, 766 of the final fixes (28.3\%) come from repository-history recovery and
would not be available from direct commit references alone.

The remaining 3,192 resolved records did not yield a verified fixing commit,
which illustrates why this recovery step is difficult.
In some, repository search finds no commit that can be confidently linked to the
vulnerability; in others, the relevant time window contains either no useful
commit history or too many plausible changes to verify reliably.
Thus, agent-assisted enrichment expands the usable evidence base, but does not
replace explicit patch references.

\PP{From fixing commits to reusable rules.}
Verification further reduces the corpus because not every security patch
provides reusable detection knowledge.
Some patches contain mostly unrelated refactoring, lack a clear vulnerable
operation, or describe conditions that are too project-specific to generalize.
After validation, 2,662 CVE cases provide usable rule evidence, from which
\sys synthesizes 1,757 rules across 266 CWE categories.
Each accepted rule retains links to its source CVEs, preserving provenance
through later consolidation and scanning.

The raw rule set is intentionally broader than the deployed knowledge base.
Closely related CWE categories frequently encode overlapping detection
conditions and anchors, which would otherwise cause the same target site to be
scanned repeatedly.
Family-level consolidation therefore merges related CWE categories and
redundant rules, reducing the store from 1,757 rules over 266 CWEs to 1,033
rules over 56 CWE families.
These rules are packaged into 172 CWE-family--language skills.
%The reduction is substantial--41.2\% of the raw rules are removed or merged---
%while the contributing CVEs remain attached to the surviving rules for
%provenance.

\PP{Coverage across languages and weakness families.}
The mined rules span both memory-safety and higher-level API-misuse patterns.
Before consolidation, the largest language-specific rule sets include Python,
PHP, Go, JavaScript, Java, TypeScript, C, and C++.
The same weakness can also appear across languages. For example, path traversal, injection,
deserialization, and related families produce language-specific rules with
different APIs but similar underlying security conditions.
Conversely, C and C++ contribute more heavily to memory-safety families.

This coverage follows the scope defined in \autoref{ss:design-challenges}.
\sys targets recurring vulnerability conditions that can be anchored at or near
security-relevant API use.
Vulnerabilities that depend primarily on global protocol state, race
conditions, or other API-independent logic may therefore have no corresponding
rule even when they appear in the CVE corpus.

%%--------------------------------------------------------------------

\subsection{Scanner Efficiency and Task Decomposition}
\label{ss:scanner-eval}

We next evaluate the two design choices behind RC2: reducing the number of
candidates that require semantic analysis and bounding the semantic work
assigned to each \phaseb worker.

\PP{Candidate reduction before semantic analysis.}
Across 15 targets, deterministic filtering reduces the aggregate \phasea
candidate pool from approximately 745K to 293K, removing 60.6\% of candidates
before model-based verification.
The benefit is largest for high-volume rules: on FreeBSD, for example,
\cc{cwe122-cpp} alone produces 65.5K call sites, which filtering reduces to
42.0K.
On smaller or lower-volume targets the reduction can be modest, supporting the
design of \phaseao as an optional cost-control layer rather than a prerequisite
for scanning.

\PP{Bootstrapping filters from \phaseb feedback.}
We further evaluate filter synthesis on 10 high-volume skills from four
targets, using up to 100 \phaseb-labeled candidates per skill (979 total).
Of these candidates, 814 are labeled false positive.
The first synthesis iteration produces conservative filters for seven skills
and removes 261 candidates, corresponding to 32.1\% of the observed false
positives; for the remaining three skills, no deterministic pattern satisfying
the safety criteria is found.
Applied to the full candidate pool, these newly synthesized filters further
reduce the workload from approximately 299K to 293K candidates.

During synthesis, candidates labeled \cc{BUGGY} or \cc{UNKNOWN} are protected:
all 165 such candidates in the synthesis samples survive the accepted filters.
This is a construction-time safety check rather than an independent recall
measurement, but it prevents the synthesis procedure from accepting filters
that contradict available positive or uncertain evidence.

\PP{Bounded semantic tasks.}
Filtering controls how many candidates reach semantic verification; task
decomposition controls how much reasoning each worker performs.
\phaseb groups only small batches from the same rule and source file, avoiding
repository-scale reasoning while retaining shared rule and source context.
We evaluate the resulting model requirement separately in
\autoref{ss:repro}, where multiple models verify the identical set of 8,115
Pillow candidates.
Together, these two mechanisms reduce both the volume and the scope of
model-based analysis.

%%--------------------------------------------------------------------
\subsection{Cross-Harness Comparison on wolfSSL}
\label{ss:harness}

\PP{Rolled-back target and five configurations}
We chose wolfSSL, an embeddable TLS/DTLS and cryptography library, because it is heavily audited and is the target of Anthropic's Mythos and Glasswing disclosure work, so its open findings give a ready basis for comparing \sys with independent harnesses.
We ran all five configurations on it under their own default settings.
The target is commit \cc{0c4ca257a07a} (wolfSSL~5.8.4), deliberately rolled back to a pre-fix state.
Two are \sys under the Pi agent with gpt-5-mini and with Kimi-K2.5, and \phasea delivers exactly the same 6{,}099 candidates to \phaseb in both.
This subsection quotes that later Pi run throughout, whereas the wolfSSL row of \autoref{t:realworld-db} quotes an earlier registry run with a smaller skill set and 3{,}166 candidates.
The three baselines are the OpenAI/Codex security-scan harness with GPT-5.5~\cite{openaicodexsec}, Anthropic's open-source \emph{defending code} reference harness with Claude Opus~4.8~\cite{anthropicdefcode}, and the Visa Vulnerability Agentic Harness (VVAH) with Claude Opus~4.8~\cite{visavvah}.

\PP{Expanded ground truth}
Recall was originally measured against nine disclosed cases from Anthropic's Mythos coordinated-disclosure program%
\footnote{\url{https://red.anthropic.com/2026/cvd/}}, and nine points cannot separate a 40\% detector from a 55\% one.
We therefore rebuilt the ground truth from the vendor release notes for 5.9.0, 5.9.1, and 5.9.2, the three releases published after the snapshot.
Those notes name 68 CVE identifiers, of which 64 localize into the rolled-back tree by aligning each fix's pre-image against the snapshot source, and the other four are excluded with a recorded reason.
A harness finding counts as a hit when it names the same file and the same enclosing function as a localized fix hunk.
We calibrated the automatic file-and-function matcher against the existing hand adjudication of the nine Mythos cases across the seven harness configurations recorded at calibration time, which gives 63 labeled decisions, and it agrees on 59 (93.7\%).
Of the four disagreements, two are false positives that credit a baseline with a case its author did not identify, so the recall gap below is conservative.

\begin{figure}[t]
\centering
\includegraphics[width=\columnwidth]{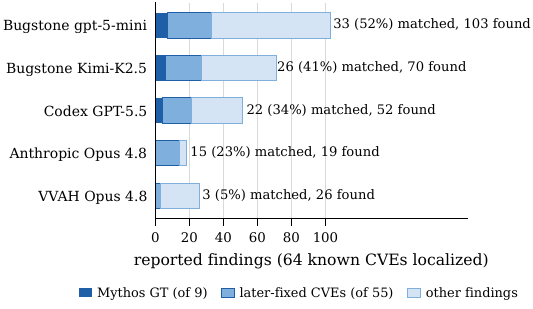}
\caption{Recall on the rolled-back wolfSSL~5.8.4 snapshot, over the 64
  post-snapshot CVEs that localize into the snapshot tree.
  Each bar spans the system's total reported findings, split into the nine
  hand-adjudicated Mythos ground-truth cases matched, the 55 later-fixed CVEs
  matched, and the remaining findings that match no disclosed CVE.
  The two matched segments therefore give the recall count, and the full bar
  gives reported volume, since the disclosed CVEs are a lower bound on true
  positives rather than all of them.
  \sys rows count \phaseb \buggy output.}
\label{f:harnessrecall}
\end{figure}

\PP{Recall on 64 cases}
\autoref{f:harnessrecall} breaks recall down by case set, and the expansion confirms the ordering the nine cases gave rather than changing it.
\sys with gpt-5-mini leads at 33 of 64 (52\%), the union of its two model configurations reaches 35 (55\%), the Codex harness reaches 22 (34\%), the Anthropic harness 15 (23\%), and VVAH 3 (5\%).
Recall on the 55 later-fixed CVEs tracks recall on all 64 within a few points for every system, so the expansion did not select a favorable population.
The expansion also corrects the earlier result in the baselines' favor.
Specifically, the Anthropic harness and VVAH score 0 of the 9 Mythos cases yet match 15 and 3 of the 64 localized CVEs, almost all of them among the 55 later-fixed ones, so reporting only the nine Mythos cases understated both.
The disclosed CVEs are a lower bound on true positives rather than the full set, since a harness can report a real bug that no vendor advisory yet covers.
Each bar spans the system's total reported findings, and the trailing count restates that volume, which the lightest segment beyond the matched cases makes visible.
The two \sys models reported 103 and 70 findings, above every baseline, since the Codex harness reported 52 and neither the Anthropic harness nor VVAH exceeded 26.

\PP{No harness subsumes another}
43 of the 64 cases are found by at least one system against 35 for the two \sys configurations together, so the baselines contribute eight cases \sys misses.
These systems are better read as an ensemble than as competitors on a single ranking.
21 of the 64 cases are found by nobody.
Four of those sit in hand-written assembly and a header that no harness here parses, so part of the residual miss set is a shared input-language limit rather than a reasoning failure.
The one Mythos case no system reports is CVE-2026-5446, ARIA-GCM explicit-IV/nonce reuse, and that miss is instructive rather than incidental.
Nonce reuse is a \emph{stateful cryptographic invariant} spanning successive record encryptions, not the misuse of a dangerous sink at one call site.
Therefore no rule in the CVE-derived rule base expresses it, which is the scope boundary already declared in \autoref{ss:cvedata}.

\PP{Caveats}
Three limits bound this comparison.
First, ground truth is vendor-disclosed CVEs only and the clone we mined ends at 2026-06-24, so silently fixed bugs and later fixes are absent.
Second, the match rule is enclosing-function identity, which is coarser than root-cause identity.
Third, this is a single C/TLS target chosen because ground truth exists for it, so the recall ordering above is not a general ranking of these harnesses.
% TODO: replicate the cross-harness comparison on a second rolled-back target with public ground truth (ideally a non-C project) to test whether the low agreement and the \sys recall margin persist.
% NOTE: t:harnesscfg, t:mythosgt, t:harnesscov and f:harnesscmp were dropped 2026-08-05.
% The stale f:harnesscmp reference in sections/openscience.tex was removed the same day,
% so no label in this list is referenced anywhere; do not resurrect one without a float.

% ---------------------------------------------------------------------------
% Appended at the end of sections/eval.tex
% Evidence root: /home/qiushi/Targets/KevinBench/multi_runs (7 Pillow runs)
% Numbers derived by fig/src/analyze_determinism.py -> determinism_stats.json
% Mechanism numbers: fig/src/variance_mechanisms.md
% ---------------------------------------------------------------------------

\subsection{Run-to-Run Reproducibility}
\label{ss:repro}

\begin{figure*}[ht]
\centering
\includegraphics[width=\textwidth]{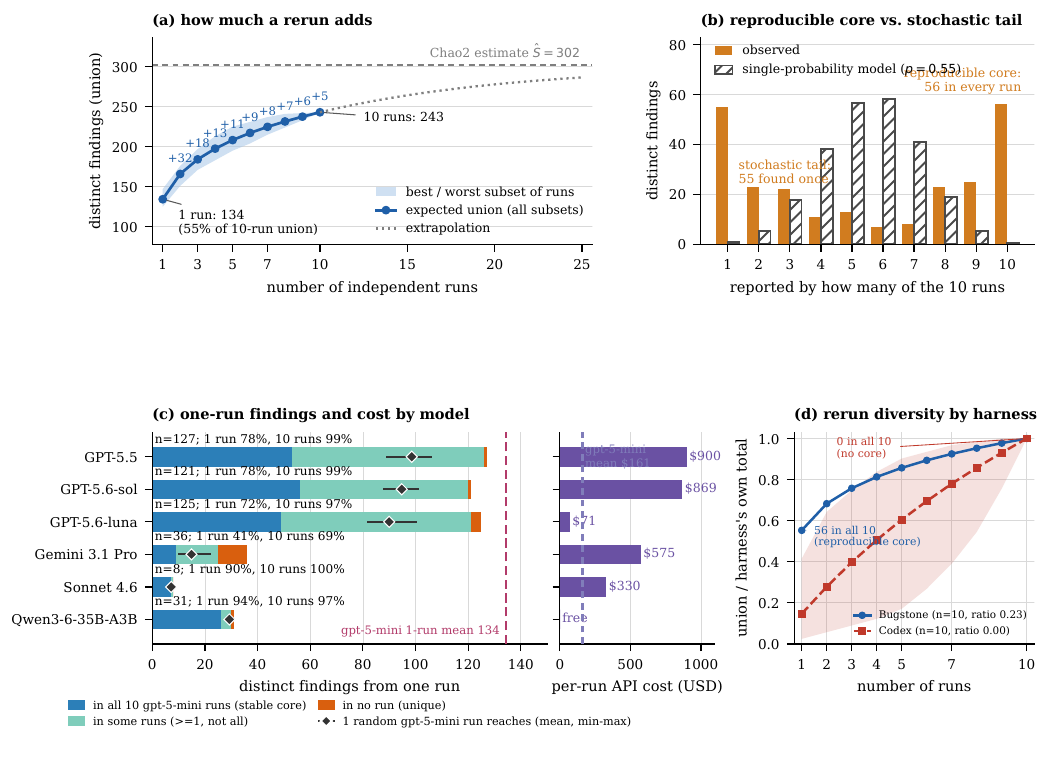}
\caption{Reproducibility of \phaseb on the same Pillow snapshot.
(a) Expected finding coverage from repeated \cc{gpt-5-mini} runs.
% with an incidence-based Chao2 extrapolation~\cite{chao1987,colwell2012} over the runs-by-locations incidence matrix.
(b) Finding frequency across ten runs. % showing a reproducible core and a stochastic tail rather than a single shared detection probability.
(c) Findings and API cost for six alternative models on the same 8,115 candidates%, with findings grouped by their overlap with the ten \cc{gpt-5-mini} runs.
(d) Growth of the finding union across reruns of \sys and the OpenAI Codex
security harness. %The harness comparison measures rerun diversity rather than like-for-like recall because the two systems report findings at different stages.
}
\label{f:repro}
\end{figure*}

\begin{table}[h]
\centering
\scriptsize
\ra{1.08}
\setlength{\tabcolsep}{3.2pt}
\begin{tabular}{lrrrrrr}
\toprule
\textbf{Run} & \textbf{Raw} & \textbf{\cc{FP}} & \textbf{Unk.}
  & \textbf{Final} & \textbf{Excl.} & \textbf{Tokens} \\
\midrule
run1  & 246 & 7{,}842 &  6 & 138 &  8 & 731M \\
run2  & 245 & 7{,}857 & 13 & 143 &  9 & 690M \\
run3  & 235 & 7{,}872 &  8 & 131 &  8 & 696M \\
run4  & 240 & 7{,}870 &  5 & 131 &  4 & 699M \\
run5  & 219 & 7{,}875 & 14 & 133 & 10 & 794M \\
run6  & 220 & 7{,}890 &  5 & 135 &  5 & 679M \\
run7  & 196 & 7{,}915 &  4 & 132 &  1 & 765M \\
run8  & 261 & 7{,}850 &  4 & 147 &  4 & 718M \\
run9  & 233 & 7{,}872 & 10 & 128 &  5 & 743M \\
run10 & 235 & 7{,}861 & 19 & 125 &  1 & 716M \\
\midrule
%\textbf{Union}        & --- & --- & --- & \textbf{243} & --- & 7.23G \\
%\textbf{Intersection} & --- & --- & --- & \textbf{56}  & --- & --- \\
\bottomrule
\end{tabular}
\caption{Ten independent runs on the same Pillow snapshot with an identical
  configuration.
  % labelled by \phaseb start order.  \phasea is deterministic and delivers the same 8{,}115 candidates in the same 1{,}345 batches to \phaseb in every run.
  \emph{Raw}: \buggy judgments before collapsing by location;
  \emph{Unk.}: \unknown verdicts.
  run1 and run5 additionally left 21 and 7 candidates unjudged, so their rows sum
  to fewer than 8{,}115.
  \emph{Final}: distinct \buggy locations;
  \emph{Excl.}: locations this run reports and no other run does.
  %Consistency ratio (intersection/union) is $0.23$; mean pairwise Jaccard similarity is $0.62$.
  }
\label{t:repro-runs}
\end{table}

\PP{Ten identical runs.}
Because \phaseb relies on an LLM agent, identical scans need not produce
identical findings.
We therefore rolled Pillow back to commit \cc{d56032047d11} and scanned it ten
times with the same runner and \cc{gpt-5-mini} model, using a frozen 174-skill
snapshot of the rule base (the current production library contains 172 skills).
All runs enumerate the same 30,594 raw matches and remove the same 21,584 with
the deterministic \phaseao filter, leaving 9,010 candidates that dedup to an
identical set of 8,115 delivered to \phaseb in the same 1,345 batches.
Any variation after this point therefore comes from semantic verification
rather than candidate delivery.

Despite identical input, individual runs report 125--147 findings
(\autoref{t:repro-runs}).
Their union contains 243 distinct locations, while only 56 appear in all ten
runs.
A typical run recovers 55.3% of this union and the best recovers 60.5%.
Thus, aggregate finding counts are relatively stable, but the identity of the
reported findings is not.

\PP{A stable core and a stochastic tail.}
The variation is highly structured rather than uniform.
Of the 243 findings in the ten-run union, 56 appear in every run and 55 appear
in only one; the remaining 132 fall between these extremes
(\autoref{f:repro}b).
At the candidate level, 93.1\% of the 8,115 candidates receive the same verdict
in all ten runs once \unknown abstentions are counted as changes; restricting to
outright \buggy--\fp flips leaves 509 contested candidates (6.3\%).
Either way, disagreement is concentrated in a relatively small set of borderline
cases.
Findings supported by multiple \buggy judgments within a run are also much more
stable across runs: 73.0\% are unanimous across all ten runs, compared with
15.8\% for findings supported by a single judgment.
This suggests that within-run corroboration provides a useful indicator of
expected stability without requiring another scan.

\PP{Reruns improve coverage with diminishing returns.}
Repeating \phaseb expands the finding set, but the marginal gain decreases
quickly.
Averaged over all subsets of the ten runs, three runs recover 75.8\% of the
ten-run union and five recover 85.7\% (\autoref{f:repro}a).
The second run adds an expected 32 findings, the third adds 18, and later runs
contribute progressively fewer.
A Chao2 incidence-based extrapolation~\cite{chao1987,colwell2012} over the
runs-by-locations incidence matrix estimates approximately 302
\phaseb-reportable locations under this harness, suggesting that even ten runs
recover only about 80.4\% of them.
This 302 is the population reachable by this rule set and candidate stage, not an
estimate of the true number of vulnerabilities in Pillow.
Repetition is therefore useful for expanding coverage, but no small number of
reruns eliminates the stochastic tail.

\PP{Where does the variation come from?}
The variation is almost entirely due to verdict changes rather than worker or
reporting failures.
Across the 1,087 missing occurrences relative to the ten-run union, 1,061
(97.6\%) are \buggy-to-\fp flips, 26 are abstentions, and none result from
worker failure.
These flips are also correlated within a semantic task: the 509 contested
candidates occur in only 153 batches, with strong within-batch correlation
($\phi=0.71$) and essentially no correlation across batches
($\phi=-0.001$).
Thus, a bounded batch behaves partly as one shared reasoning decision, rather
than as several independent candidate classifications.
This observation motivates keeping \phaseb batches small even when candidates
share rule and file context.

\PP{Reproducibility is not recall.}
More findings from repeated runs do not necessarily imply proportionally higher
vulnerability recall.
The rolled-back Pillow snapshot contains 21 vulnerabilities disclosed and fixed
after the selected revision, but only five reach \phaseb as candidates.
Across ten runs, \phaseb reports four of those five at least once, while a
typical run reports only one.
The remaining sixteen are lost before \phaseb: fourteen are never enumerated at
all, eight of them because the deployed skill set has no rule for their CWE, and
two more are enumerated but removed by the \phaseao filter.
Repetition can therefore recover semantic-verification misses, but it cannot
compensate for missing detection knowledge or candidate coverage.

\PP{Rerunning a small model versus using a stronger model.}
We next run six additional models once each on the same Pillow snapshot and the
same 8,115 \phaseb candidates.
A single \cc{gpt-5-mini} run already covers 78\% of the findings reported by
GPT-5.5, 78\% of GPT-5.6-sol, and 72\% of GPT-5.6-luna, close to the 79\%
overlap between two \cc{gpt-5-mini} runs.
Across ten \cc{gpt-5-mini} runs, coverage of these stronger-model finding sets
rises to 97--99\%, leaving at most four findings unique to any one model
(\autoref{f:repro}c).

Several frontier configurations are also substantially more expensive
(\autoref{t:model-cost}).
One \cc{gpt-5-mini} run averages approximately \$161, compared with about
\$900 for GPT-5.5 and \$869 for GPT-5.6-sol on the same candidate set, while
reporting a comparable or larger number of \phaseb findings.
For this bounded verification task, spending the same budget on several small
model runs therefore provides greater \phaseb finding coverage than replacing the
small model with a single frontier-model run.
This result supports the RC2 design choice: once semantic analysis is
decomposed into narrow candidate-level tasks, \phaseb does not require the
strongest available model for every judgment.

\PP{Harness dependence and limitations.}
The benefit of rerunning is not universal to all agentic scanners.
Ten runs of the OpenAI Codex security harness on the same Pillow snapshot show
substantially less recurring overlap than \sys
(\autoref{f:repro}d), indicating that rerun behavior depends on how a harness
structures its reasoning tasks.
We treat this comparison as a diversity observation rather than a
like-for-like recall result because the harnesses expose findings at different
stages.

Finally, this study covers one target and one primary \phaseb configuration,
and the repeated runs stop before dynamic validation.
The additional findings in the stochastic tail therefore should not be
interpreted as confirmed vulnerabilities.
Instead, the experiment measures the stability and coverage behavior of the
semantic-verification stage itself.

\begin{table}[t]
\centering
\scriptsize
\ra{1.08}
\setlength{\tabcolsep}{4.2pt}
\begin{tabular}{@{}lrrrr@{}}
\toprule
\textbf{Model} & \textbf{\$/1M in/out} & \textbf{Tokens}
  & \textbf{Findings} & \textbf{Cost} \\
\midrule
\cc{gpt-5-mini}   & 0.25 / 2.00 & 723M & 134 & \$161 \\
\cc{gpt-5.6-luna} & 0.20 / 1.20 & 479M & 125 & \$71 \\
\cc{sonnet-4.6}   & 3.00 / 15.0 & 166M &   8 & \$330 \\
\cc{gemini-3.1-pro} & 2.00 / 12.0 & 311M & 36 & \$575 \\
\cc{gpt-5.6-sol}  & 5.00 / 30.0 & 311M & 121 & \$869 \\
\cc{gpt-5.5}      & 5.00 / 30.0 & 309M & 127 & \$900 \\
\cc{qwen3-6-35b-a3b} & self-hosted & 346M & 31 & free \\
\bottomrule
\end{tabular}
\caption{Per-run \phaseb cost on the same Pillow snapshot and candidates.
  \emph{\$/1M}: listed input and output prices per million tokens.
  \emph{Cost} is provider-billed for the whole run and includes retries and
  cached-input tokens priced well below the listed input rate, so it is not the
  product of the \emph{Tokens} column and the in/out rates.
  \cc{gpt-5-mini} \emph{Findings} and \emph{Cost} are the ten-run mean; the
  other rows are a single run.  Prices are the provider list rates as of
  2026-08.
  \emph{free}: Qwen3-6-35B-A3B runs on a local A100, so it incurs no per-token
  API charge. }
\label{t:model-cost}
\end{table}

% ---------------------------------------------------------------------------
% Phase D evaluation.
% Evidence root: /home/qiushi/PaperWritting/staging_phased/phased_data_inventory.md
% Derived from 5 real Phase D runs under /home/qiushi/BugStoneSkills/results/
% and /home/qiushi/BugStoneDocker/results/.  The 71 other phase_d/ directories
% in those trees are synthetic no-LLM test fixtures and are excluded.
% ---------------------------------------------------------------------------

\subsection{Remediation Prototype}
\label{ss:phased}

Patch quality is difficult to evaluate automatically: whether a repair is
complete, regression-free, and acceptable to maintainers often depends on
project-specific tests and human review.  We therefore evaluate \phased only as
a prototype for exploit-guided remediation rather than as a general patch
correctness system.

We apply \phased to 23 findings across 3 targets for which \phasec
produces a reproducible proof of concept.  Only one of these targets,
the openai-python SDK (11 of the 23 findings), is an independent third-party
codebase; the other two are a small internal Flask exercise and a C++
vulnerability-hunting exercise.  For each finding, \phased generates
a patch and applies the two-sided differential test from
\autoref{ss:scanner}: the proof of concept must reproduce on the original
program, stop reproducing after the patch, and reproduce again after the patch
is reverted.  \phased produced a patch for all 23 findings, and all 23 pass this
differential check.  The resulting patches are compact, with a median of 8
changed lines (6 added, 2 removed), and 21 of the 23 touch a single file.

These results provide machine-checkable evidence that the current prototype can
turn a live finding into a compact candidate remediation that blocks the
demonstrated behavior.  The differential outcome is the worker's own transcribed
runtime output rather than an execution we re-ran, and this small sample does not
measure general patch correctness or maintainer acceptance, which require
substantially broader project-specific testing and review.

\section{Real-World Bug Detection}
\label{s:realworld}

\sys scanned 14 third-party projects and recorded every finding in a deduplicating registry.
\autoref{t:realworld-db} reports the outcome per target, from the \phasea candidate pool through the \phasec verification tiers.
Each row is one project, and the finding counts are deduplicated across every configuration run on it, so a finding seen by two configurations is counted once.
The registry omits Pillow, which is scanned only for the reproducibility study of \autoref{ss:repro}.

% GENERATED by staging_fixes/vulndb/derive.py --latex-table (tier columns),
% then hand-merged: the codex-src and claude-code rows and the Lang/Domain/Cand
% target-metadata columns are added on top of derive.py's 12-project output.
% Source of record for finding counts: ~/BugStoneVulnerabilityDB.
% All 14 rows count toward the totals.  Pillow carries no registry finding
% counts (reproducibility-only) and is commented out, so its cells would be "--".
% WordPress is reported at its true 18 (registry stores 14; five stored-XSS
% reports share one file+line and collapse to one record -- the $\ddagger$ note).
\begin{table*}[t]
\centering
\scriptsize
\ra{1.08}
\setlength{\tabcolsep}{4pt}

\label{t:realworld-db}
\begin{tabular}{@{}lllrrrrrr@{}}
\toprule
\textbf{Target} & \textbf{Lang.} & \textbf{Domain} & \textbf{Cand.}
  & \textbf{\phaseb} & \textbf{\phasec} & \textbf{Runtime} & \textbf{Live} & \textbf{Conf.} \\
  %& \textbf{Role} \\
\midrule
FreeBSD                  & C/C++       & OS kernel      & 76{,}843 & 2{,}189 & 1{,}485 & 459 & 40 & 16 \\%& O \\
PyTorch                  & Python      & ML framework   & 38{,}605 & 295 & 195 & 95 & 95 & 27\\% & C \\
OpenSSL                  & C/C++       & Crypto lib.    & 20{,}357 & 20 & 16 & 5 & 5 & 2 \\% & C \\
%Pillow                   & Python/C    & Imaging lib.   & 9{,}010  & -- & -- & -- & -- & -- \\%  & R \\
WordPress$^{\ddagger}$   & PHP/JS      & CMS            & 8{,}636  & 18 & 18 & 0 & 0 & 0  \\% & C \\
NVIDIA OGK               & C           & GPU driver     & 6{,}449  & 26 & 26 & 8 & 8 & 0 \\% & O \\
ImageMagick              & C/C++       & Image tooling  & 5{,}161  & 162 & 162 & 4 & 4 & 2   \\% & O \\
wxo-clients              & Python      & Enterprise SDK & 3{,}907  & 46 & 46 & 41 & 41 & 0 \\%& C \\
guava                    & Java        & Java core lib. & 3{,}843  & 22 & 22 & 0 & 0 & 0 \\%& C \\
openssh                  & C           & SSH impl.      & 3{,}441  & 1 & 1 & 0 & 0 & 0 \\%& C \\
wolfSSL$^{\dagger}$      & C           & Embedded TLS   & 3{,}166  & 31 & 31 & 16 & 16 & 4 \\% & H \\
openai-python            & Python      & API client     & 841      & 34 & 34 & 0 & 0 & 0 \\% & O \\
MCP servers              & TS/JS       & Agent servers  & 739      & 36 & 36 & 12 & 12 & 10\\ % & O \\
%\midrule
codex-src                & Python/Rust & Coding agent   & 1{,}150  & 47 & 47 & 2 & 2 & 0 \\% & O, V \\  
claude-code              & Python/TS   & Coding agent   & 586      & 6 & 6 & 2 & 1 & 1 \\% & O, V \\  
\midrule
\textbf{Total} & & & & \textbf{2{,}933} & \textbf{2{,}125} & \textbf{644} & \textbf{224} & \textbf{62} \\% & \\  

\bottomrule
\end{tabular}
\caption{Scanned targets, ordered by candidate volume, with deduplicated
  finding counts from the \sys registry.
  \emph{Cand.}\ is \phasea candidates after the \phaseao filter; \emph{\phaseb}
  distinct \buggy locations; \emph{\phasec} findings reaching \phasec;
  \emph{Runtime} a sanitizer, crash, or PoC signal at the sink; \emph{Live} an
  exploitability of \cc{LIVE\_POC}; and \emph{Conf.}\ the strongest \confexp tier.
  Each tier is strictly stronger than the one to its left.
  $^{\dagger}$wolfSSL is the cross-harness target of \autoref{ss:harness}; this
  row quotes an earlier registry run with a smaller skill set (3{,}166
  candidates) rather than the later rolled-back Pi run analyzed there.
  $^{\ddagger}$Five WordPress stored-XSS reports share one file and line and
  collapse to a single registry record, so the row is reported at its true 18.
  %\emph{Role}: \textbf{C} controlled multi-configuration comparison, \textbf{H}
  %cross-harness comparison (\autoref{ss:harness}), \textbf{R} reproducibility
  %study (\autoref{ss:repro}), \textbf{O} opportunistic single run, \textbf{V}
  %vendor coding-agent self-analysis excluded from registry totals.
  %Pillow (\textbf{R}) and the two \textbf{V} coding agents are outside the
  %registry, so their finding cells are ``--'' and the headline count is 12
  %third-party projects.
  %$^{\dagger}$wolfSSL is the rolled-back cross-harness target of \autoref{ss:harness},
  %and this row quotes the registry run rather than the later Pi run.
  %$^{\ddagger}$five WordPress stored-XSS reports collapse to one registry record.
  }
\end{table*}

\PP{The scan portfolio by verification tier}
The columns of \autoref{t:realworld-db} are strictly nested, so a reader can pick an evidentiary bar and read off the count.
Every \confexp finding has a live proof-of-concept, and every live proof-of-concept is a runtime observation.
Across the 14 targets, \sys deduplicated 2{,}933 \phaseb findings, of which \phasec adjudicated 2{,}125 and 644 carry runtime evidence, meaning a sanitizer report, a crash, or a controlled-sink signal at the reported location.
Runtime evidence is the headline bar we report, and the stricter live-proof-of-concept and \confexp tiers in \autoref{t:realworld-db} narrow it further for readers who want them.
Of the 2{,}125 findings \phasec adjudicated, it rejected 1{,}099 as not a bug or not exploitable and left 382 with a static-reachability argument but no runtime signal, so 644 reached the runtime bar.
The gap between the \phaseb and \phasec columns is a separate matter: it is backlog, not a negative result.
Specifically, 808 findings have never been given a runtime at all, 704 of them in FreeBSD.
Reporting the rejected set and the untested backlog separately is what keeps ``not exploitable'' distinct from ``could not be tested,'' so the runtime-evidenced counts are lower bounds set by how much \phasec compute we spent, not by how many findings survive scrutiny.

\PP{Phase A enumeration}
Summed across the 14 targets, the \emph{Cand.}\ column of \autoref{t:realworld-db} totals 173{,}724 post-filter candidates that reach \phaseb, after the deterministic \phaseao filter has already removed benign sites without any model call.
Each candidate then reaches \phaseb, which follows the callers, weighs every guard on the taint path, and returns a \buggy verdict only with code evidence.

\PP{Phase B and Phase C findings}
Runtime-evidence rates differ sharply by target and by weakness class.
On guava, all 22 \phaseb deserialization candidates reach \phasec, but none yet carries runtime evidence, so guava contributes no findings at the runtime bar.
The single openssh finding, a CWE-190 integer overflow in \cc{setenv.c:183}, illustrates precision in a widely audited codebase, since 3{,}441 candidates narrow to exactly one \phaseb finding, still unverified dynamically.
PyTorch concentrates on RCE-class deserialization bugs, with 195 findings reaching \phasec and 95 carrying a runtime observation.
FreeBSD dominates raw volume, with 2{,}189 \phaseb findings against a 76{,}843-candidate pool.
Only 459 have a runtime signal so far, because the \phasec backlog there is the largest.

\PP{False positives cluster in three shapes}
Dominant false-positive sources fall into three categories.
\emph{Benign wrappers} are internal utility functions that call a dangerous API with a constant or already-validated argument, where the \phasea pattern matches but \phaseb identifies the effective guard.
\emph{Machine-generated code} is the second category.
For example, openssh's libcrux\_mlkem768 ML-KEM implementation contains $\approx$635 syntactic buffer-write matches, and all were correctly dismissed because the generated code uses bounded index arithmetic throughout.
\emph{Incomplete taint chains} are the third category, in which the dangerous sink exists but user-controlled input does not reach it within the analyzed call depth.
\phaseb labels these \fp with a taint-chain explanation, so the structured evidence format enables principled labeling rather than a bare verdict.

\PP{CWE families and languages}
Runtime-evidenced findings concentrate in families with deep rule coverage.
CWE-502 insecure deserialization dominates PyTorch and guava, CWE-79 cross-site scripting dominates WordPress, and the C memory-safety families CWE-120, CWE-122, CWE-125, CWE-787, and CWE-416 account for the OpenSSL and FreeBSD findings.
CWE-22 path traversal, CWE-77 and CWE-78 command injection, and CWE-918 server-side request forgery carry runtime evidence in wxo-clients under multiple configurations.
The runtime-evidenced findings therefore span injection and deserialization as well as memory-safety families, and they cover six languages from C to TypeScript.
This spread matches the cross-language coverage of the rule base in \autoref{ss:cvedata}.

\PP{The scan model shifts volume, not runtime evidence}
The scan model changes how many candidates \phaseb marks \buggy, but \phasec then filters that raw volume down to what a runtime can support.
Under the controlled Pillow comparison of \autoref{ss:repro}, larger models and gpt-5-mini report comparable \phaseb finding counts on the same candidate set, so a stronger scan model mainly reshapes which candidates are flagged rather than how many survive runtime checking.
The extra candidates a noisier model raises are not free: each one still consumes \phasec compute, and the 808-finding \phasec backlog above is exactly the cost of that raw volume.
Splitting \phaseb into a simple per-candidate judgment keeps each judgment cheap enough for a small model, while the dynamic \phasec confirmation carries the load-bearing runtime evidence.
Therefore \sys does not depend on the strongest and most expensive model to produce runtime-evidenced findings.
A harness that concentrates all reasoning in one large model has the opposite property.

\section{Discussion}
\label{s:discuss}

\subsection{Limitations}
\label{ss:limitations}

\PP{Detection scope and completeness}
\sys is designed for recurring vulnerability patterns that can be anchored to security-relevant operations or API uses.  It therefore does not target one-off design flaws, configuration errors, or global and stateful invariants without a suitable local anchor.  CVE-derived rules may also remain incomplete. A rule can miss semantically equivalent APIs or conditions not represented in the source fixes.  In addition, conservative candidate filtering and bounded \phaseb reasoning can introduce false negatives, particularly when a decision requires deeper interprocedural context.  Because \phaseb is model-driven, its verdicts are also stochastic; reruns can recover some semantic-analysis misses but cannot recover vulnerabilities outside the deployed rule and candidate coverage.

\PP{Runtime validation}
Runtime confirmation depends on the target admitting a reproducible build and test environment.  Complex dependencies, unavailable inputs, or platform-specific behavior can prevent \phasec from exercising an otherwise plausible finding.  Such cases should therefore be interpreted as unverified rather than benign.  Dynamic validation is also necessarily more expensive than static verification, so large Phase-B finding sets may only be partially exercised.

\PP{Remediation guarantees}
\phased remains a prototype.  Its differential oracle establishes that a patch
blocks the demonstrated proof of concept and that reverting the patch restores
the behavior.  It does not establish that the repair is complete,
regression-free, or acceptable to maintainers.  Evaluating these properties at
scale requires project-specific testing and substantial human review, so we
treat remediation as an end-to-end capability rather than a measured patch
correctness result.

\subsection{Future Work}
\label{ss:futurework}

Future work can extend \sys along these limitations.  Broader CVE coverage and
rule auditing can improve generality across CWE families, while richer
interprocedural context can increase detection coverage beyond locally anchored
cases.  Runtime validation can likewise benefit from more reusable build
environments and broader confirmation of \phaseb findings.  Remediation
requires substantially more project-specific and human-in-the-loop evaluation:
future versions of \phased can incorporate native regression tests, assess
whether patches address the complete root cause, and track maintainer feedback
as more patches are submitted.  We therefore view the current remediation
stage as an evolving prototype rather than a complete automated repair system.

\section{Related Work}
\label{s:relwk}

\PP{Static and query-based vulnerability analysis}
Code property graphs, QL and CodeQL, Semgrep, and Infer enumerate candidates from human-authored queries or rules~\cite{yamaguchi2014modeling,avgustinov2016ql,codeql,semgrep,calcagno2015moving,bennett2024semgrepstar}.  Recent work attacks that authoring bottleneck.  Specifically, QLCoder synthesizes CodeQL queries from CVE metadata, and QRS generates queries agentically and validates findings by exploit synthesis~\cite{wang2025qlcoder,tsigkourakos2026qrs}.  SemTaint likewise extracts per-package taint specifications for CodeQL~\cite{ghebremichael2026semtaint}.  Therefore \sys is not the first system to turn CVE history into executable detectors.  However, \sys emits CWE- and language-level rules that group verified fixing commits, keep \cc{source\_cves} provenance, and carry an agent-consumed verification contract.

\PP{Vulnerability clone and patch-signature detection}
ReDeBug, VUDDY, MVP, MOVERY, V1SCAN, and VMud reuse vulnerability history through code, component, or patch-line signatures~\cite{jang2012redebug,kim2017vuddy,xiao2020mvp,woo2022movery,woo2023v1scan,huang2024vmud}.  MAVM replaces those signatures with agents that detect, confirm, repair, and validate recurring cases~\cite{zheng2026mavm}.  In contrast, \sys matches CWE- and API-level rules, so a target need not clone the seed vulnerability, and confirmation requires an executed exploit.

\PP{Patch analysis and rule inference}
Empirical studies and datasets such as BigVul show that security patches encode security properties~\cite{li2017securitypatches,fan2020bigvul}.  Prior work infers those properties for missing checks, security impact, disordered error handling, and severity prioritization~\cite{lu2019crix,wu2020symradar,wu2021eh,wu2022diffcvss}.  Closer to \sys, VulGenie derives Java API rules by attack-defense cross-analysis, while RuleForge and RulePilot convert CVE records and analyst annotations into validated detection rules~\cite{chen2026vulgenie,garg2026ruleforge,wang2025rulepilot}.  In addition, AutoTrace turns one fixing commit into a trigger location, and GONDAR supplies CWE-specific sink knowledge to exploit agents~\cite{zibaeirad2026autotrace,fleischer2026gondar}.  However, \sys differs in the unit and scale of reuse, since it groups public CVE history into CWE-language rules packaged as deployable agent skills.

\PP{Learning-based vulnerability prediction}
VulDeePecker, SySeVR, Devign, ReVeal, LineVul, DeepDFA, and FVD-DPM learn vulnerability signals from gadgets, graphs, or transformer encodings~\cite{li2018vuldeepecker,li2022sysevr,zhou2019devign,chakraborty2021reveal,fu2022linevul,steenhoek2024deepdfa,shao2024fvddpm}.  That line now extends into agent training.  For example, VulAgentRL rewards only verdicts whose evidence checks against a code property graph, and Antares distills compact localization models~\cite{li2026vulagentrl,vijay2026antares}.  However, these systems predict over learned representations, whereas \sys keeps CVE provenance per rule and demands taint, guard, and caller rationale.

\PP{LLMs for vulnerability detection}
LLift, Vul-RAG, IRIS, LLMxCPG, and specialized reasoning models guide LLMs with static analysis, graphs, or retrieved vulnerability knowledge~\cite{li2024llift,du2024vulrag,li2025iris,lekssays2025llmxcpg,nie2025vulnllmr}.  Benchmark studies measure that behavior across model sizes, prompts, repositories, and agent settings~\cite{lin2025mammoth,sun2024llm4vuln,yildiz2025benchmarking,nie2025secodeplt}.  Repository-scale agents followed quickly.  Specifically, TitanCA reports a deployed match, filter, inspect, and adapt pipeline, LLMVD.js exploits taint bugs in Node.js packages, and DREA separates exploration from reasoning~\cite{zhang2026titanca,ni2026llmvdjs,sun2026drea}.  Evaluation moved with it, since VulnGym, SastBench, and RealVuln supply traces, triage distributions, and scanner rankings~\cite{ji2026vulngym,feiglin2026sastbench,pellew2026realvuln}.  Also, agentic filtering removes most SAST noise yet suppresses true positives, and a Vul-RAG replication reports an accuracy plateau~\cite{xiong2026sifting,kaniewski2026revisiting}.  Therefore \sys layers filtering into a cost-ordered cascade, in which \phaseao rejects deterministically and only survivors reach an agent.

\PP{Agentic program repair and self-validating patches}
AgenticRepair assembles repair context through subagents, while KeaRepair grounds patches in verified facts and mined vulnerability--patch pairs~\cite{fu2026agenticrepair,cao2026kearepair}.  However, acceptance criteria are fragile, since adversarial issue reports yield newly vulnerable patches and most Java CVE patches fail stricter oracles~\cite{chen2025swexploit,almaamari2026whyllmsfail}.  Execution-based patch acceptance therefore has prior art.  Specifically, VulnRepairEval requires the original exploit to fail, Vul4Py adds a functional oracle, and VeriPort chains evidence across affected versions~\cite{wang2025vulnrepaireval,bui2026vul4py,ghebremichael2026veriport}.  In contrast, \phased patches findings that \sys discovered itself, so acceptance needs a two-sided test in which the exploit fails under the patch and succeeds after reverting it.

\PP{Exploit generation, dynamic confirmation, and agent determinism}
Big Sleep, ATLANTIS, and the AIxCC systematization show agents finding and repairing real bugs alongside fuzzing and symbolic execution~\cite{bigsleep2024naptime,kim2025atlantis,zhang2026aixcc}.  K-Repro, SEC-bench Pro, and CVE-Bench measure proof-of-concept construction on kernel N-days, browser engines, and web CVEs~\cite{pu2026patchtopoc,lee2026secbenchpro,zhu2025cvebench}.  However, agents still fail on service-based vulnerabilities whose environments they cannot stand up~\cite{liu2025webrepro}.  RECEIPT restores verdict trust through isolation and binding, and FalseCrashReducer validates caller context against infeasible crashes~\cite{lyu2026receipt,amusuo2025falsecrashreducer}.  Repeated-run studies report that single-run scores overstate coverage, that inference backends shift results, and that injected static structure halves variance~\cite{zhou2026stability,pape2026backends,lin2026anchoring}.  Therefore \sys fixes its \phasea candidate set before any model runs and measures run-to-run reproducibility, so execution collapses the remaining spread.

\PP{Security taxonomies and public vulnerability data}
\sys builds on public infrastructure, namely \cc{cvelistV5} records, NVD metadata, the CWE taxonomy, and OSV advisories~\cite{cvelistv5,nvd,cwe,osv}.  CVEfixes, MoreFixes, MegaVul, and VulZoo unify CVE-linked fixes, while ThreatKG and VulnScopper mine missing relations among security entities~\cite{bhandari2021cvefixes,akhoundali2024morefixes,ni2024megavul,ruan2024vulzoo,shi2024threatkg,alfasi2024vulnscopper}.  Therefore \sys differs by turning verified fixing commits into executable scanner inputs with provenance rather than into datasets.

\section{Conclusion}
\label{s:conclusion}

\sys turns vulnerability history into reusable detection knowledge and applies
it through a staged workflow from scalable detection to runtime validation and
candidate remediation.  Its current knowledge base contains 1,033 rules
consolidated into 56 CWE families and deployed through 172 skills.  Across
controlled experiments and real-world scans of 14 projects, these skills
surface recurring vulnerability conditions beyond the projects from which they
were originally derived, with 644 findings supported by runtime evidence.
More broadly, \sys concentrates increasingly expensive analysis on
progressively fewer candidates while grounding stronger security claims in
observable evidence rather than model judgment alone.  Our results show that
historical vulnerability fixes can serve not only as records of past failures,
but also as reusable knowledge for finding and validating future ones.

% The two mandated USENIX appendices come *before* the bibliography, per the
% official template (usenixsecurity2026.tex:248-261).  Their titles are fixed
% strings -- see the header comment in each file.  Any *optional* appendix goes
% after Open Science, so a reviewer meets the mandated two first.
%\cleardoublepage

%\input{sections/ethics}
%\input{sections/openscience}
%\cleardoublepage
% Optional appendix: full-detail tables pulled out of the body to meet the
% 13-page body limit.  Must stay after Open Science.

%\cleardoublepage
\bibliographystyle{abbrvnat}
% conf.bib must come first: it defines the venue @STRING macros p.bib entries use.
\bibliography{conf,p}
%\appendix
%\input{sections/appendix}
\end{document}